\documentclass[
superscriptaddress,
 amsmath,amssymb,
 aps,
 prc,
 twocolumn,
floatfix,
longbibliography
]{revtex4-2}
\usepackage{graphicx}  
\usepackage{dcolumn}   
\usepackage{bm}        
\usepackage{amssymb}   
\usepackage{hyperref}
\usepackage{color}
\usepackage[utf8]{inputenc}
\usepackage{orcidlink}
\usepackage{braket}
\usepackage{CJK}

\begin{document}
\begin{CJK}{UTF8}{gkai}

\title{Electric dipole polarizability of $^{58}$Ni}\thanks{Notice: This manuscript has been authored, in part, by UT-Battelle, LLC, under contract DE-AC05-00OR22725 with the US Department of Energy (DOE). The US government retains and the publisher, by accepting the article for publication, acknowledges that the US government retains a nonexclusive, paid-up, irrevocable, worldwide license to publish or reproduce the published form of this manuscript, or allow others to do so, for US government purposes. DOE will provide public access to these results of federally sponsored research in accordance with the DOE Public Access Plan (https://www.energy.gov/doe-public-access-plan).}

\newcommand{\TUDarmstadt}{Institut f\"ur Kernphysik, Technische Universit\"at Darmstadt, D-64289 Darmstadt, Germany}
\newcommand{\UM}{Institut f\"ur Kernphysik and PRISMA+ Cluster of Excellence, Johannes Gutenberg-Universit\"at Mainz, D-55128 Mainz, Germany}
\newcommand{\FRIB}{Facility for Rare Isotope Beams, Michigan State University, East Lansing, MI 48824, USA}
\newcommand{\RCNP}{Research Center for Nuclear Physics, Osaka University, Ibaraki, Osaka 567-0047, Japan}
\newcommand{\Erlangen}{Institut f\"ur Theoretische Physik II, Universit\"at Erlangen, D-91058 Erlangen, Germany}
\newcommand{\Milanoa}{Dipartimento di Fisica “Aldo Pontremoli”, Universit\`a degli Studi di Milano, 20133 Milano, Italy}
\newcommand{\Milanob}{INFN, Sezione di Milano, 20133 Milano, Italy}
\newcommand{\FHU}{Faculty of Radiological Technology, Fujita Health University, Aichi 470-1192, Japan}
\newcommand{\ORNLone}{Physics Division, Oak Ridge National Laboratory,
Oak Ridge, TN 37831, USA} 
\newcommand{\ORNLtwo}{National Center for Computational Sciences, Oak Ridge National Laboratory,
Oak Ridge, TN 37831, USA} 
\newcommand{\Barcelonaa}{Departament de F\'isica Qu\`antica i  F\'isica, Mart\'i i Franqu\'es, 1, 08028 Barcelona, Spain}
\newcommand{\Barcelonab}{Institut de Ci\`encies del Cosmos, Universitat de Barcelona, Mart\'i i Franqu\'es, 1, 08028 Barcelona, Spain}
\newcommand{\Peking}{State Key Laboratory of Nuclear Physics and Technology, School of Physics, Peking University, Beijing 100871, China}
\newcommand{\Lanzhoua}{School of Nuclear Science and Technology, Lanzhou University, Lanzhou 730000, China}
\newcommand{\Lanzhoub}{Frontiers Science Center for Rare isotope, Lanzhou University, Lanzhou 730000, China}

\author{I.~Brandherm}\affiliation{\TUDarmstadt}
\author{F.~Bonaiti}\affiliation{\UM}\affiliation{\FRIB}\affiliation{\ORNLone}
\author{P.~von~Neumann-Cosel}\email[Email:]{vnc@ikp.tu-darmstadt.de}\affiliation{\TUDarmstadt}
\author{S.~Bacca}\affiliation{\UM}
\author{G.~Col\`o}\affiliation{\Milanoa}\affiliation{\Milanob}
\author{G.~R.~Jansen}\affiliation{\ORNLtwo}\affiliation{\ORNLone}
\author{Z.~Z.~Li (李征征)}\affiliation{\Peking}\affiliation{\Lanzhoua}\affiliation{\Lanzhoub}
\author{H.~Matsubara}\affiliation{\RCNP}\affiliation{\FHU}
\author{Y.~F.~Niu (牛一斐)}\affiliation{\Lanzhoua}\affiliation{\Lanzhoub}
\author{P.-G.~Reinhard}\affiliation{\Erlangen}
\author{A.~Richter}\affiliation{\TUDarmstadt}
\author{X.~Roca-Maza}\affiliation{\Barcelonaa}\affiliation{\Barcelonab}\affiliation{\Milanoa}\affiliation{\Milanob}
\author{A.~Tamii}\affiliation{\RCNP}

\begin{abstract}
The electric dipole strength distribution in $^{58}$Ni between 6 and 20~MeV has been determined  from proton inelastic scattering experiments at very forward angles at RCNP, Osaka.
The experimental data are rather well reproduced by quasiparticle random-phase approximation calculations including vibration coupling, despite a mild dependence on the adopted Skyrme interaction.
They allow an estimate of the  experimentally inaccessible high-energy contribution above 20 MeV, leading to an electric dipole polarizability $\alpha_\mathrm{D}(^{58}{\rm Ni}) = 3.48(31)$~fm$^3$.
This serves as a test case for recent extensions of coupled-cluster calculations with chiral effective field theory interactions to nuclei with two nucleons on top of a closed-shell system. 

\end{abstract}

\date{\today}

\maketitle 

\end{CJK}

{\em Introduction}.--
The nuclear equation of state (EOS) governs basic properties of nuclei~\cite{roc18} and neutron stars \cite{lat21,hut22} as well as the dynamics of core-collapse supernovae \cite{yas20} and neutron star mergers~\cite{raa21}. 
A systematic description of the EOS from nuclear densities to those in neutron stars is a central goal of current physics. 
A wealth of new data is available at high densities from observations on the properties of neutron stars and neutron star mergers but the present experimental constraints on the EOS around the saturation density $n_0$ of nuclear matter are still insufficient.

The EOS of symmetric nuclear matter is rather well constrained \cite{roc18} in contrast to the properties of neutron-rich matter.
The latter depends on the symmetry energy, which can be  parameterized in an expansion around $n_0$ by the symmetry energy at saturation density $J(n_0)$ and its density dependence  $L = 3 n_0 \partial J(n_0)/ \partial n$.
Higher-order terms are expected to be small.
There are many experimental methods \cite{thi19} providing constraints on $J$ and $L$ based on 
a model-dependent correlation between $L$ and the neutron-skin thickness $r_{\rm skin}$ in nuclei with neutron excess~\cite{bro00,cen09,rei10,roc11}. 
For a recent summary, see Ref.~\cite{lat23}.
The electric dipole polarizability, $\alpha_{\rm D}$, has also been identified as a key observable for constraining EOS parameters~\cite{rei10,roc13}. 
Proton inelastic scattering at incident energies of several hundred MeV at extreme forward angles 
has been developed as a new experimental tool exactly for the study of $\alpha_{\rm D}$ \cite{vnc19a} and results have been provided for a wide range of nuclei~\cite{tam11,has15,bir17,bas20a,fea23}.

Two theoretical approaches have been used to describe $\alpha_{\rm D}$ and derive constraints on the symmetry energy parameters: energy density functional theory (DFT)~\cite{ben03,piekarewicz2012,roc18} and {\it ab initio} calculations~\cite{hag16,sim19,hu21} starting from chiral two- and three-nucleon interactions~\cite{heb11,eks15}.
A correlation of the form $\alpha_{\rm D} \cdot J \propto L$, suggested by the droplet model, has been well studied in DFT~\cite{roc13,roc15}. In the {\it ab initio} context, comparing experimental determinations of $\alpha_D$ with theoretical predictions allows to validate constraints on nuclear matter properties from chiral forces.

In such efforts, coupled-cluster (CC) theory~\cite{hag16} plays a prominent role. Successful comparisons between CC predictions and experimental data in $^{40,48}$Ca~\cite{bir17,fea23} and $^{68}$Ni~\cite{kau20} have established this approach as an ideal tool to describe $\alpha_D$ in closed-shell, medium-mass nuclei.
The same method has also been applied at the dripline to study the low-energy dipole strength and polarizability of $^8$He~\cite{bonaiti2022,bonaiti2024_fbs}. Very recently, the reach of coupled-cluster calculations of $\alpha_D$ has been extended beyond closed-shell nuclei~\cite{bonaiti2024}. This new development has focused on two-particle-attached (2PA) systems, characterized by two nucleons outside a closed-shell nucleus. Combining closed-shell and 2PA coupled-cluster predictions, Ref.~\cite{bonaiti2024} enabled an analysis of the evolution of the dipole polarizability along the oxygen and calcium isotopic chains. 

In this Letter, we present a measurement of the dipole polarizability of $^{58}$Ni. Having two neutrons outside the doubly magic $^{56}$Ni, it serves as a test case for the newly developed 2PA method. A study of $^{58}$Ni is also of interest to systematically explore the theoretically predicted dependence on neutron skin thickness when combined with data for $^{64}$Ni (presently under analysis) and $^{68}$Ni \cite{ros13}.

{\em Experiment}.-- 
The $^{58}$Ni$(p,p')$ reaction has been measured at RCNP, Osaka, at an incident proton energy of 295 MeV in a laboratory scattering angle range $0.4^\circ - 5.15^\circ$ and for excitation energies in the range $5 - 25$ MeV.
An energy resolution of 22 keV (full width at half maximum) was achieved applying dispersion matching techniques.
The experimental techniques and the raw data analysis are described in Ref.~\cite{tam09}.
Further details of the $^{58}$Ni experiment are described in Ref.~\cite{bra24} presenting a state-by-state analysis of electric and magnetic dipole transitions at excitation energies up to 13 MeV. 
These data provide information on the isovector spinflip $M1$ resonance and candidates for a toroidal $E1$ mode in nuclei \cite{vnc23}. 

\begin{figure}
\includegraphics[width=\columnwidth]{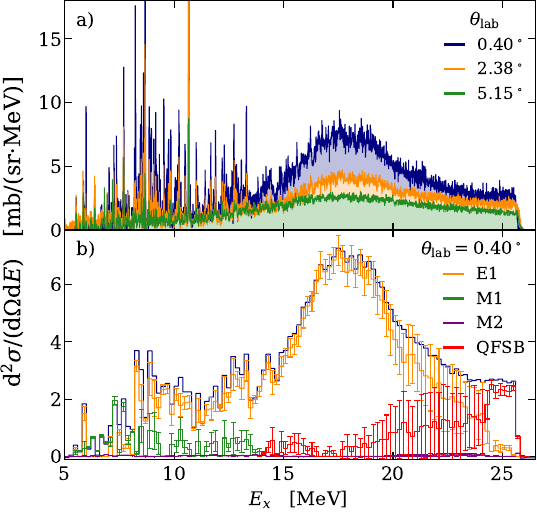}
\caption{(a) Spectra of the $^{58}$Ni$(p,p')$ reaction at $E_0 = 295$~MeV and scattering angles $\Theta_{\rm lab} = 0.40^\circ$, $2.38^\circ$ and $5.15^\circ$.
(b) Example of the MDA of the spectrum at $\Theta_{\rm lab} = 0.4^\circ$ in 200 keV bins (blue) and decomposition into contributions of $E1$ (orange), $M1$ (green), multipoles $\lambda > 1$ (purple), and an empirical background (red). 
Contributions from the ISGMR and ISGQR were subtracted prior to the MDA as described in the text.
\label{fig:spectra}}
\end{figure}

The top panel of Fig.~\ref{fig:spectra} presents representative energy spectra measured at laboratory scattering angles $\Theta_{\rm lab} = 0.40^\circ$, $2.38^\circ$, and $5.15^\circ$. 
The cross sections above 10 MeV show a broad resonance with a maximum at about 18 MeV  and cross sections strongly decreasing with scattering angle.
The angular dependence is consistent with relativistic Coulomb excitation of $E1$ transitions.
Thus, we identify this resonance structure as the IsoVector Giant Dipole Resonance (IVGDR).

The various contributions to the spectra were separated using a multipole decomposition analysis (MDA) as described e.g.\ in Ref.~\cite{bas20b}.
Theoretical angular distributions for the relevant multipoles were obtained from distorted wave Born approximation calculations with transition amplitudes from quasiparticle-phonon-model calculations, cf.~Ref.~\cite{bra24}. 
Since only spectra at seven angles were available, the number of multipoles that can be considered in the MDA was limited.
Following the method described in Refs.~\cite{bas20b,don18} and using the experimental $E0$ and $E2$ strength distributions in $^{58}$Ni from inelastic $\alpha$ scattering \cite{liu06}, the contributions to the spectra due to excitation of the isoscalar giant monopole and quadrupole resonances were subtracted prior to the MDA. 
Additionally, an empirical background (most likely due to quasifree scattering) was considered. 
Its angular dependence was taken from experiments on heavier nuclei \cite{tam11,bas20b}, which showed a momentum-transfer dependence approximately independent of nuclear mass.

Results for the most forward angle measured are presented in the bottom part of Fig.~\ref{fig:spectra} as example, where the spectra was rebinned to 200 keV.
$E1$ cross sections dominate over the whole excitation energy range.
At energies up to 13.5 MeV, the spinflip $M1$ resonance makes sizable contributions.
The two strongest $M1$ transitions (cf.~Ref.~\cite{bra24}) were subtracted by hand because they lead to large uncertainties in the MDA for the respective energy bins.
The background becomes relevant on the high-energy flank of the IVGDR.
Above 20 MeV, the uncertainties of the $E1$/background decomposition become very large due to the similarity of their angular distributions.
Contributions from higher multipoles are negligibly small.

\begin{figure}
\includegraphics[width=\columnwidth]{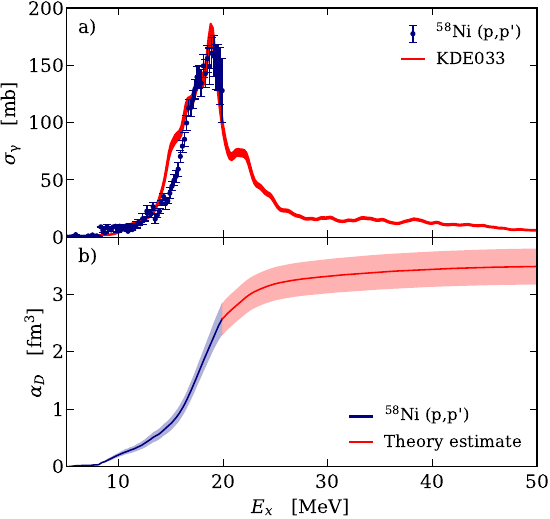}
\caption{
(a) Photoabsorption cross sections of $^{58}$Ni derived from the spectrum at a scattering angle of $0.40^\circ$ using the virtual photon method (blue circles).
The red curve shows a QRPA calculation including qPVC \cite{li23} with the KDE033 interaction \cite{agr05} normalized to the data. 
(b) Electric dipole polarizability $\alpha_D$ derived from the photoabsorption cross sections. 
The blue and red bands band show the present data and the contribution at excitation energies $> 20$ MeV based on the theoretical estimate explained in the text with their uncertainties, respectively.
\label{fig:absalpha}}
\end{figure}

{\em Extraction of the dipole polarizability}.--
The $E1$ cross sections resulting from the MDA were converted into equivalent photoabsorption cross sections using the virtual photon method~\cite{ber88}.  
The virtual photon spectrum was calculated in an eikonal approach \cite{ber93} to Coulomb excitation, integrated over the distribution of scattering angles covered in the solid angle of each angular bin as described in Ref.~\cite{bas20b}. 
The resulting photoabsorption cross sections are displayed as blue circles in Fig.~\ref{fig:absalpha}(a).

The electric dipole polarizability $\alpha_D$ is related to the photoabsorption cross sections by 
\begin{equation}
\label{eq:pol}
  \alpha_\mathrm{D} = \frac{\hbar c}{2\pi^{2} } 
  \int \frac{\sigma_\mathrm{\gamma}}{E_{\rm x}^{2}}{\rm d}E_{\rm x}.
\end{equation}
The experimental result for the energy region $6 - 20$ MeV is plotted as blue curve in Fig.~\ref{fig:absalpha}(b) and amounts to $\alpha_{\rm D} = 2.57(28)$ fm$^3$.
The uncertainty band considers the systematic errors of the experimental cross sections (cf.~Ref.~\cite{bra24}) and the MDA (as described in Ref.~\cite{bas20b}).
Statistical uncertainties are negligible.

Photoabsorption data from the ($\gamma,xn)$ reaction are available for excitation energies up to 33 MeV \cite{ful74}, but in contrast to heavy nuclei the unknown $(\gamma,p)$ channel is expected to be significant. 
Thus, $\alpha_{\rm D}$ contributions at energies $E_{\rm x} > 20$ MeV were estimated with a theory-aided procedure using energy density functionals.
Previous analyses of this type \cite{bas20a,ros13} were based on the folding of QRPA calculations with interactions reproducing the IVGDR centroid with a Lorentzian fitted to the experimental data.
Here, we go beyond and include quasiparticle vibration coupling (qPVC) which has recently been shown to permit not only a reproduction of the width of the ISGMR \cite{lit23,li23}, but also resolve the discrepancies between $^{208}$Pb and lighter nuclei in theoretical attempts to extract the compressibility from the energy centroid of the ISGMR \cite{gar18}.

\begin{figure}
\includegraphics[width=\columnwidth]{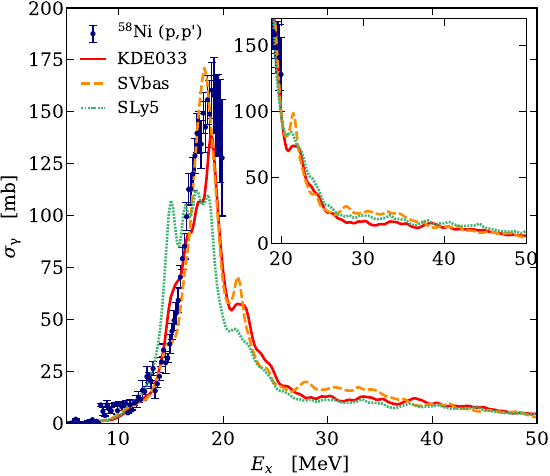}
\caption{
Photoabsorption cross sections of $^{58}$Ni from the present work compared with QRPA calculations including qPVC \cite{li23}, based on the KDE033 (solid red line) \cite{agr05}, SV-bas (dashed orange line) \cite{klu09}, and SLy5 (short-dashed green line) \cite{cha97} interactions.  
The inset shows the high-energy flanks normalized to each other at 20 MeV.
\label{fig:highenergy}
}
\end{figure}

QRPA calculations including qPVC with the approach described in Ref.~\cite{li23} are shown in Fig.~\ref{fig:highenergy} for Skyrme forces KDE033 \cite{agr05}, SV-bas \cite{klu09}, and SLy5 \cite{cha97}.
The photoabsorption cross sections predicted with KDE033 (solid red line) provide a very good desription of the centroid and width of the IVGDR, but the total strength is somewhat underestimated.
Calculations with SV-bas (dashed orange line) give a similar width and reproduce the maximum cross section, but the centroid energy is about 1 MeV too low.
Finally, the SLy5 result (short-dashed green line) shows a much stronger fragmentation and an even lower energy centroid.
Since all calculations require an adjustment to the data, the absolute values of the different models for the high-energy ($> 20$ MeV) contribution to the polarizability becomes very dependent on the assumptions made in the normalization procedure.

For a quantitative estimate of the high-energy contribution to the polarizability, we choose a normalization to the results obtained with the KDE033 interaction.
As illustrated in Fig.~\ref{fig:absalpha}(a), it provides a very good description of the IVGDR after adjusting the absolute height. 
The corresponding contribution to the polarizability for excitation energies $> 20$~MeV is displayed in Fig.~\ref{fig:absalpha}(b) as red curve.
The polarizability is integrated up to 50 MeV, where saturation is reached.

The model dependence due to the choice of specific interactions is estimated from the variation of the three calculations after normalization to each other at 20 MeV.
As demonstrated in the inset of Fig.~\ref{fig:highenergy}, then the theoretically predicted high-energy tails become similar in shape and magnitude.
The similar energy dependence might look surprising at first sight but can be understood from the following argument: structures on the low-energy side of the IVGDR are related to the coupling to individual collective phonons, which leads to the phenomenon of fine structure \cite{pol14,car22}.
At higher excitation energies stochastic coupling \cite{vnc19b} predominates, i.e., the strength distribution is mainly determined by the density of states and an average coupling matrix element between the particle-hole and more complex states.  

The theoretically predicted contribution to the polarizability amounts to $\alpha_{\rm D}(E_{\rm x} > 20 \, {\rm MeV}) = 0.91$ fm$^3$ with an uncertainty of 0.04 fm$^3$ due to the normalization.
As pointed above, a model-dependent error is estimated from the variation of the three calculations (0.13 fm$^3$).
The parameter dependence of the individual calculations for the different forces was estimated from variations of the cutoff energy of the single-particle spectrum and the minimum strength of phonons considered in the qPVC and found to be negligibly small.
Assuming that all above-discussed error contributions are independent, we find for the dipole polarizability $\alpha_{\rm D}(^{58}{\rm Ni}) = 3.48 (31)$ fm$^3$.

{\em Coupled-cluster calculations}.--
The electric dipole polarizability of Eq.~(\ref{eq:pol}) is a sum rule of  the photoabsoption cross section, which can be itself written in terms of the dipole response function
\begin{equation}
   R(E_{\rm x}) =  \sum_{\mu} |\braket{\Psi_{\mu}|\Theta|\Psi_0}|^2 \delta(E_{\mu} - E_0 - E_{\rm x}) \,,
\label{response}
\end{equation}
where $\ket{\Psi_0}$ and $\ket{\Psi_{\mu}}$  are ground and excited state of the nucleus with energy $E_0$ and $E_{\mu}$, respectively, while $\Theta$ is the dipole operator. 
The sum over $\mu$ in Eq. (\ref{response}) runs over bound and continuum excited states of the nucleus. This makes the calculation of response functions particularly challenging, because of the presence of unbound configurations arising from the break-up of the nucleus into fragments. To avoid this, one can resort to the Lorentz Integral Transform (LIT) method~\cite{efros2007}, which is based on an integral transform of the response function with a Lorentzian kernel  as
\begin{equation}
    L(\sigma,\Gamma) = \frac{\Gamma}{\pi} \int dE_{\rm x}\; \frac{R(E_{\rm x})}{(E_{\rm x}-\sigma)^2 + \Gamma^2}.
\label{transform}
\end{equation}
Calculating the latter requires ``only" the solution of a bound-state problem.
 Because the Lorentzian kernel tends to a Dirac delta function as $\Gamma\rightarrow 0$, one has that
\begin{equation}
    L(\sigma, \Gamma\rightarrow 0) = \int dE_{\rm x}\;R(E_{\rm x})\delta(E_{\rm x} -\sigma) = R(\sigma)\,,
\label{litsmallgamma}
\end{equation}
which effectively means that in this limit the LIT becomes the response function, where the variable $E_{\rm x}$ is renamed to $\sigma$. Such response function is discretized in the sense that excited states in the continuum are represented by bound pseudo-states. 
Nevertheless, Eq. (\ref{litsmallgamma}) can be used  to compute the $n-$th moments  from the response function as
\begin{equation}
   m_n =   \int dE_{\rm x}\; E_{\rm x}^n\; R(E_{\rm x}) = \int d\sigma\; \sigma^n L(\sigma, \Gamma\rightarrow 0)\,.
\label{mn_def}
\end{equation}
Given that sum rules can be written as expectation values on the ground-state, the utilization of bound pseudo-states in such a calculation is mathematically valid~\cite{NevoDinur:2014ngu}. With this reasoning, the electric dipole polarizabilty is simply related to the inverse energy-weighted sum rule of the dipole response function as $\alpha_{\rm D} = 2\alpha \hbar c m_{-1}$, where $ m_{-1}$ is calculated using Eq.~(\ref{mn_def}), and $\alpha$ is the fine structure constant.

Merging the LIT approach~\cite{efros2007} with the coupled-cluster theory~\cite{hagen2014} for closed-(sub) shell nuclei led to a method dubbed LIT-CC, which is based on the following steps~\cite{bac13}.  First, the ground state  is constructed starting from a Slater determinant  ($\Phi_0$) and imprinting correlations on top of it via an exponential ansatz  $\ket{\Psi_0} = e^{T} \ket{\Phi_0}$.
The cluster operator $T$ can be expanded in terms of a sum of $n$-particle $n$-hole excitations. Second, the Hamiltonian and the excitation operator are similarity transformed to $\overline{H} = e^{-T} H e^T$  and $\overline{\Theta} = e^{-T} \Theta e^T$, respectively.  
Third, excited states of a  closed-(sub) shell nucleus are computed as $\ket{\Psi_{\mu}} = R_{\mu} e^T \ket{\Phi_0}$, where the operator $R_\mu$ is also expanded in terms of particle-hole excitations, by solving an equation of motion. Finally, $\alpha_D$ can be computed with the prescription described above.

While the LIT-CC method has been  very successfully used for closed-shell nuclei \cite{bac13,hag16,bir17,mio18,sim19,bonaiti2022,fea23}, recently it has been extended to nuclei that have two nucleon on top of a closed shell system using the two-particle attached (2PA) technique~\cite{bonaiti2024}. In this case, excited states of 2PA nuclei can be obtained with the following ansatz
\begin{equation}
    \ket{\Psi_{\mu}^{\rm(A+2)}} = R^{A+2}_{\mu} \ket{\Psi_0^{(A)}} = R^{A+2}_{\mu} e^T \ket{\Phi_0^{(A)}},
    \label{2pa-ansatz}
\end{equation}
where  $A$ is the mass number of the closed-(sub)shell system and the excitation operator $R^{A+2}_{\mu}$ involves the net creation of the two extra nucleons on top of the closed-(sub)shell system
\begin{equation}
    R^{A+2}_{\mu} = \frac{1}{2} \sum_{ab} r^{ab}_{\mu} a^{\dagger}_a a^{\dagger}_b + \frac{1}{6} \sum_{abci} r^{abc}_{i, \mu} a^{\dagger}_a a^{\dagger}_b a^{\dagger}_c a_i\,.
    \label{rmu-2pa}
\end{equation}
In this work we adopt the particle-hole expansion of Eq.~(\ref{rmu-2pa}), including two-particle zero-hole (2p-0h) and three-particle one-hole (3p-1h) contributions. The addition of higher-order terms is at the moment prohibitive. In Ref.~\cite{bonaiti2024}, this method was employed to study $\alpha_D$ in oxygen and calcium isotopes. Here, we apply it to a larger mass number by studying the $^{58}$Ni nucleus, starting from the $^{56}$Ni closed-shell neighbour. 

We perform our calculation using the chiral nucleon-nucleon and three-nucleon interactions 1.8/2.0 (EM)~\cite{heb11}, $\Delta$NLO$_{\rm GO}(450)$ and $\Delta$NNLO$_{\rm GO}(450)$~\cite{jiang20}. The chiral force 1.8/2.0 (EM) yields accurate binding energies in medium-mass and heavy nuclei~\cite{sim17}, and it contains two-nucleon forces up to next-to-next-to-next-to-leading order, softened via similarity renormalization group transformation at a scale of $1.8$ fm$^{-1}$, and three-nucleon forces at next-to-next-to-leading order, with a momentum cutoff of $2.0$ fm$^{-1}$. The interactions $\Delta$NLO$_{\rm GO}(450)$ and $\Delta$NNLO$_{\rm GO}(450)$ contain the $\Delta$-isobar as an explicit degree of freedom, and they are given at next-to-leading order and next-to-next-to-leading order, respectively. 

Our 2PA calculations of $\alpha_D$ start from an Hartree-Fock reference state, expanded on a harmonic oscillator basis of up to $13$ major shells. We studied the convergence of our results varying the underlying harmonic oscillator frequency $\hbar\Omega$ between $12$ and $16$ MeV. An additional energy cut at $E_{3,max} = 16\hbar\Omega$ is applied on three-body contributions.  

\begin{figure}
\includegraphics[width=\columnwidth]{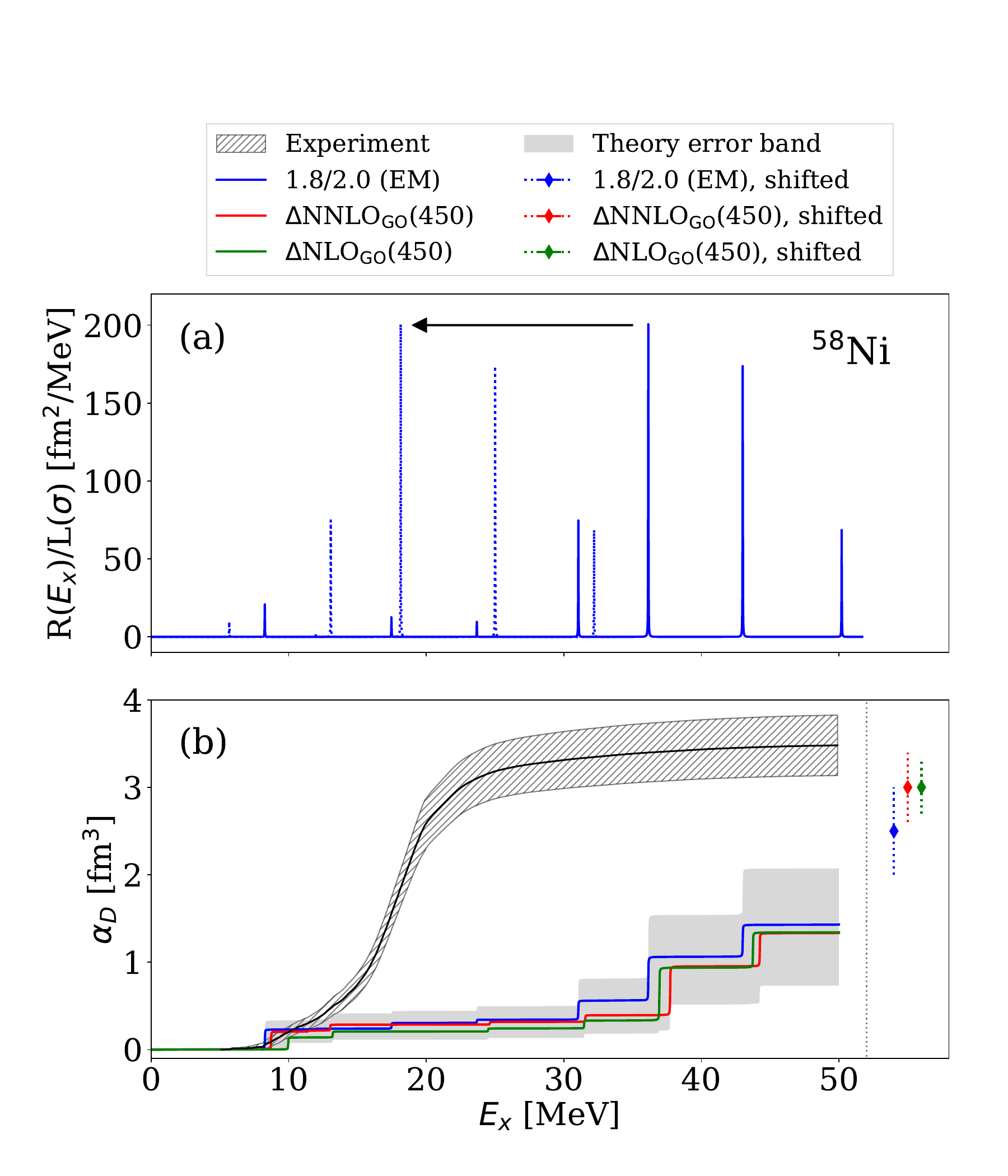}
\caption{
(a) Discretized response function of $^{58}$Ni calculated with the 1.8/2.0 (EM) interaction (solid line), and the corresponding curve obtained shifting the response to the experimental IVGDR energy of 18 MeV (dotted line). 
(b) $\alpha_{\rm D}$ running sums for the 1.8/2.0 (EM) and $\Delta$-full interaction models, in comparison to experiment. 
$\alpha_{\rm D}$ predictions obtained by shifting the response to the experimental IVGDR energy are shown on the r.h.s.\ as diamonds with dotted error bars. 
\label{fig:theory}
}
\end{figure}

In Figure~\ref{fig:theory}, in the upper panel we show the discretized response function  versus  the excitation energy calculated with the  1.8/2.0 (EM) interaction. We clearly see one peak located below $10$ MeV, while the four largest peaks which constitute the IVGDR are found at high energy, beyond $30$ MeV. In the lower panel we compare instead the theoretical and experimental running sum rules, defined as in Eq.~(\ref{eq:pol}), where the integral is performed up to a maximum upper limit, which is varied from 0 to 50 MeV. The experimental data are shown with a hatched band, while in case of the theoretical results, we present three different interactions: the 1.8/2.0 (EM) in blue, the $\Delta$NLO$_{\rm GO}(450)$ and $\Delta$NNLO$_{\rm GO}(450)$  in green and red, respectively.

Interestingly, we see that experiment and theory agree with each other at low energy. Considering the interaction dependence and the many-body truncation error, estimated according to the recipe devised in Ref.~\cite{bonaiti2024}, we find $0.1< \alpha_D < 0.3$ fm$^3$ below 11 MeV of excitation energy, in good accordance with the corresponding experimental result of $0.2< \alpha_D < 0.3$ fm$^3$ obtained from the MDA analysis. At higher energies, however, the rise of the experimental running sum rule is much faster than that of the theoretical  calculations. The reason for this behaviour lies in the fact that the IVGDR pseudo-states are found at higher energies with respect to the experiment, approximately 20 MeV too high. We have already observed such effect in other isotope chains~\cite{bonaiti2024} and it is most likely related with the truncation of Eq.~(\ref{rmu-2pa}) at the 3p-1h level, which does not grasp all the necessary correlations. To gauge the possible role of missing higher-order correlations, we can compare the share of the pseudo-states' norm in 2p-0h configurations to the corresponding total norm. As a rule of thumb, if the 2p-0h contribution to the norm is around $90\%$, the nuclear state of interest has a simple 2PA structure and an accurate description of it can be achieved employing the 3p-1h approximation~\cite{jansen2011,jansen2013}. In the case of $^{58}$Ni, the 2p-0h contribution to the total norm is above $70\%$ for the first excited states at around 10 MeV, where the theoretical running sum agrees with experiment. At higher energies, instead, it falls quickly below $50\%$, suggesting the need of higher order contributions to the 2PA expansion. This is reflected by the extent of the theoretical error band shown in grey. It contains the many-body truncation uncertainty, which is clearly dominating over the potential dependence indicated by the difference of the three coloured curves. 

In order to check that the major rise of the experimental running sum is given by the IVGDR states, we took the theoretical discretized response functions and shifted all peaks of around 20 MeV, so that for each interaction the largest peak is located at the same energy as the experimental IVGDR. By doing so, $\alpha_D$ is clearly enhanced to a value compatible with the experimental result, as shown by the dotted error bars at the right of the lower panel. 
\\
{\em Conclusions}.--
We have determined the electric dipole polarizability of $^{58}$Ni from 300 MeV inelastic proton scattering experiments at very forward angles.
The extraction is limited to an excitation energy of 20 MeV due to the quasifree continuum background which shows a similar angular distribution as the Coulomb excitation cross sections.
QRPA calculations including qPVC provide a good description of the energy centroid and width of the IVGDR in $^{58}$Ni, permitting an estimate of the experimentally inaccessible strength above 20 MeV. 

The resulting value of $\alpha_{\rm D}$ serves as a test case for {\it ab initio}-based coupled cluster calculations with the newly developed 2PA method to describe nuclei with two particles outside closed-shell systems. 
At the present level of the CC expansion the low-energy $E1$ strength can be predicted well, but the IVGDR is found at an excitation energy about 20 MeV too high resulting in correspondingly small theoretical $\alpha_{\rm D}$ values. 
This points to the need for higher order many-body correlations in the 2PA expansion in order to achieve a more accurate description of the IVGDR contribution to the polarizability. To address this issue, alternative approaches could be also pursued in the future, such as coupling the LIT method to coupled-cluster calculations employing axially-symmetric reference states~\cite{novario2020,hagen2024,sun2024,hu2024}.

\begin{acknowledgments}
This work was supported by the Deutsche Forschungsgemeinschaft (DFG, German Research Foundation) -- Project-ID 279384907 -- SFB 1245 and through the Cluster of Excellence ``Precision Physics, Fundamental Interactions, and Structure of Matter" (PRISMA$^+$ EXC 2118/1, Project ID 39083149), and by the U.S. Department of Energy, Office of Science, Office of Advanced Scientific Computing Research and Office of Nuclear Physics, Scientific Discovery through Advanced Computing (SciDAC) program (SciDAC-5 NUCLEI) and Office of Nuclear Physics, under the FRIB Theory Alliance award DE-SC0013617. This research used resources of the Oak Ridge Leadership Computing Facility located at Oak Ridge National Laboratory, which is supported by the Office of Science of the Department of Energy under contract No.\ DE-AC05-00OR22725.
Computer time was provided by the Innovative and Novel Computational Impact on Theory and Experiment (INCITE) program and by the supercomputer Mogon at Johannes Gutenberg Universit\"at Mainz. 
X.R.-M.\ acknowledges support by Grant No.\ PID2020-118758GB-I00 funded by MCIN/AEI/10.13039/501100011033; by the "Unit of Excellence Mar\'ia de Maeztu 2020-2023" award to the Institute of Cosmos Sciences, Grant No.\ CEX2019-000918-M funded by MCIN/AEI/10.13039/501100011033; and by the Generalitat de Catalunya, Grant No.\ 2021SGR01095.
Y.F.N.\ acknowledges funding from National Key Research and Development (R\&D) Program under Grant No.\ 2021YFA1601500 and Natural Science Foundation of China under Grant No.\ 12075104.
P.-G.R.\ acknowledges supply of computing resources from the
RRZE of the Friedrich-Alexander university Erlangen/N\"urnberg.

\end{acknowledgments}

\bibliography{58Ni-DP}

\begin{thebibliography}{62}%
\makeatletter
\providecommand \@ifxundefined [1]{%
 \@ifx{#1\undefined}
}%
\providecommand \@ifnum [1]{%
 \ifnum #1\expandafter \@firstoftwo
 \else \expandafter \@secondoftwo
 \fi
}%
\providecommand \@ifx [1]{%
 \ifx #1\expandafter \@firstoftwo
 \else \expandafter \@secondoftwo
 \fi
}%
\providecommand \natexlab [1]{#1}%
\providecommand \enquote  [1]{``#1''}%
\providecommand \bibnamefont  [1]{#1}%
\providecommand \bibfnamefont [1]{#1}%
\providecommand \citenamefont [1]{#1}%
\providecommand \href@noop [0]{\@secondoftwo}%
\providecommand \href [0]{\begingroup \@sanitize@url \@href}%
\providecommand \@href[1]{\@@startlink{#1}\@@href}%
\providecommand \@@href[1]{\endgroup#1\@@endlink}%
\providecommand \@sanitize@url [0]{\catcode `\\12\catcode `\$12\catcode
  `\&12\catcode `\#12\catcode `\^12\catcode `\_12\catcode `\%12\relax}%
\providecommand \@@startlink[1]{}%
\providecommand \@@endlink[0]{}%
\providecommand \url  [0]{\begingroup\@sanitize@url \@url }%
\providecommand \@url [1]{\endgroup\@href {#1}{\urlprefix }}%
\providecommand \urlprefix  [0]{URL }%
\providecommand \Eprint [0]{\href }%
\providecommand \doibase [0]{https://doi.org/}%
\providecommand \selectlanguage [0]{\@gobble}%
\providecommand \bibinfo  [0]{\@secondoftwo}%
\providecommand \bibfield  [0]{\@secondoftwo}%
\providecommand \translation [1]{[#1]}%
\providecommand \BibitemOpen [0]{}%
\providecommand \bibitemStop [0]{}%
\providecommand \bibitemNoStop [0]{.\EOS\space}%
\providecommand \EOS [0]{\spacefactor3000\relax}%
\providecommand \BibitemShut  [1]{\csname bibitem#1\endcsname}%
\let\auto@bib@innerbib\@empty
\bibitem [{\citenamefont {Roca-Maza}\ and\ \citenamefont {Paar}(2018)}]{roc18}%
  \BibitemOpen
  \bibfield  {author} {\bibinfo {author} {\bibfnamefont {X.}~\bibnamefont
  {Roca-Maza}}\ and\ \bibinfo {author} {\bibfnamefont {N.}~\bibnamefont
  {Paar}},\ }\bibfield  {title} {\bibinfo {title} {Nuclear equation of state
  from ground and collective excited state properties of nuclei},\ }\href
  {https://doi.org/https://doi.org/10.1016/j.ppnp.2018.04.001} {\bibfield
  {journal} {\bibinfo  {journal} {Prog. Part. Nucl. Phys.}\ }\textbf {\bibinfo
  {volume} {101}},\ \bibinfo {pages} {96} (\bibinfo {year} {2018})}\BibitemShut
  {NoStop}%
\bibitem [{\citenamefont {Lattimer}(2021)}]{lat21}%
  \BibitemOpen
  \bibfield  {author} {\bibinfo {author} {\bibfnamefont {J.}~\bibnamefont
  {Lattimer}},\ }\bibfield  {title} {\bibinfo {title} {Neutron stars and the
  nuclear matter equation of state},\ }\href
  {https://doi.org/https://doi.org/10.1146/annurev-nucl-102419-124827}
  {\bibfield  {journal} {\bibinfo  {journal} {Annu. Rev. Nucl. Part. Sci.}\
  }\textbf {\bibinfo {volume} {71}},\ \bibinfo {pages} {433} (\bibinfo {year}
  {2021})}\BibitemShut {NoStop}%
\bibitem [{\citenamefont {Huth}\ \emph {et~al.}(2022)\citenamefont {Huth},
  \citenamefont {Pang}, \citenamefont {Tews}, \citenamefont {Dietrich},
  \citenamefont {Le~F{\`e}vre}, \citenamefont {Schwenk}, \citenamefont
  {Trautmann}, \citenamefont {Agarwal}, \citenamefont {Bulla}, \citenamefont
  {Coughlin},\ and\ \citenamefont {Van Den~Broeck}}]{hut22}%
  \BibitemOpen
  \bibfield  {author} {\bibinfo {author} {\bibfnamefont {S.}~\bibnamefont
  {Huth}}, \bibinfo {author} {\bibfnamefont {P.~T.~H.}\ \bibnamefont {Pang}},
  \bibinfo {author} {\bibfnamefont {I.}~\bibnamefont {Tews}}, \bibinfo {author}
  {\bibfnamefont {T.}~\bibnamefont {Dietrich}}, \bibinfo {author}
  {\bibfnamefont {A.}~\bibnamefont {Le~F{\`e}vre}}, \bibinfo {author}
  {\bibfnamefont {A.}~\bibnamefont {Schwenk}}, \bibinfo {author} {\bibfnamefont
  {W.}~\bibnamefont {Trautmann}}, \bibinfo {author} {\bibfnamefont
  {K.}~\bibnamefont {Agarwal}}, \bibinfo {author} {\bibfnamefont
  {M.}~\bibnamefont {Bulla}}, \bibinfo {author} {\bibfnamefont {M.~W.}\
  \bibnamefont {Coughlin}},\ and\ \bibinfo {author} {\bibfnamefont
  {C.}~\bibnamefont {Van Den~Broeck}},\ }\bibfield  {title} {\bibinfo {title}
  {Constraining neutron-star matter with microscopic and macroscopic
  collisions},\ }\href {https://doi.org/10.1038/s41586-022-04750-w} {\bibfield
  {journal} {\bibinfo  {journal} {Nature}\ }\textbf {\bibinfo {volume} {606}},\
  \bibinfo {pages} {276} (\bibinfo {year} {2022})}\BibitemShut {NoStop}%
\bibitem [{\citenamefont {Yasin}\ \emph {et~al.}(2020)\citenamefont {Yasin},
  \citenamefont {Sch\"afer}, \citenamefont {Arcones},\ and\ \citenamefont
  {Schwenk}}]{yas20}%
  \BibitemOpen
  \bibfield  {author} {\bibinfo {author} {\bibfnamefont {H.}~\bibnamefont
  {Yasin}}, \bibinfo {author} {\bibfnamefont {S.}~\bibnamefont {Sch\"afer}},
  \bibinfo {author} {\bibfnamefont {A.}~\bibnamefont {Arcones}},\ and\ \bibinfo
  {author} {\bibfnamefont {A.}~\bibnamefont {Schwenk}},\ }\bibfield  {title}
  {\bibinfo {title} {Equation of state effects in core-collapse supernovae},\
  }\href {https://doi.org/10.1103/PhysRevLett.124.092701} {\bibfield  {journal}
  {\bibinfo  {journal} {Phys. Rev. Lett.}\ }\textbf {\bibinfo {volume} {124}},\
  \bibinfo {pages} {092701} (\bibinfo {year} {2020})}\BibitemShut {NoStop}%
\bibitem [{\citenamefont {Raaijmakers}\ \emph {et~al.}(2021)\citenamefont
  {Raaijmakers}, \citenamefont {Greif}, \citenamefont {Hebeler}, \citenamefont
  {Hinderer}, \citenamefont {Nissanke}, \citenamefont {Schwenk}, \citenamefont
  {Riley}, \citenamefont {Watts}, \citenamefont {Lattimer},\ and\ \citenamefont
  {Ho}}]{raa21}%
  \BibitemOpen
  \bibfield  {author} {\bibinfo {author} {\bibfnamefont {G.}~\bibnamefont
  {Raaijmakers}}, \bibinfo {author} {\bibfnamefont {S.~K.}\ \bibnamefont
  {Greif}}, \bibinfo {author} {\bibfnamefont {K.}~\bibnamefont {Hebeler}},
  \bibinfo {author} {\bibfnamefont {T.}~\bibnamefont {Hinderer}}, \bibinfo
  {author} {\bibfnamefont {S.}~\bibnamefont {Nissanke}}, \bibinfo {author}
  {\bibfnamefont {A.}~\bibnamefont {Schwenk}}, \bibinfo {author} {\bibfnamefont
  {T.~E.}\ \bibnamefont {Riley}}, \bibinfo {author} {\bibfnamefont {A.~L.}\
  \bibnamefont {Watts}}, \bibinfo {author} {\bibfnamefont {J.~M.}\ \bibnamefont
  {Lattimer}},\ and\ \bibinfo {author} {\bibfnamefont {W.~C.~G.}\ \bibnamefont
  {Ho}},\ }\bibfield  {title} {\bibinfo {title} {{Constraints on the Dense
  Matter Equation of State and Neutron Star Properties from NICER's Mass-Radius
  Estimate of PSR J0740+6620 and Multimessenger Observations}},\ }\href
  {https://doi.org/10.3847/2041-8213/ac089a} {\bibfield  {journal} {\bibinfo
  {journal} {Astrophys. J. Lett.}\ }\textbf {\bibinfo {volume} {918}},\
  \bibinfo {pages} {L29} (\bibinfo {year} {2021})}\BibitemShut {NoStop}%
\bibitem [{\citenamefont {Thiel}\ \emph {et~al.}(2019)\citenamefont {Thiel},
  \citenamefont {Sfienti}, \citenamefont {Piekarewicz}, \citenamefont
  {Horowitz},\ and\ \citenamefont {Vanderhaeghen}}]{thi19}%
  \BibitemOpen
  \bibfield  {author} {\bibinfo {author} {\bibfnamefont {M.}~\bibnamefont
  {Thiel}}, \bibinfo {author} {\bibfnamefont {C.}~\bibnamefont {Sfienti}},
  \bibinfo {author} {\bibfnamefont {J.}~\bibnamefont {Piekarewicz}}, \bibinfo
  {author} {\bibfnamefont {C.~J.}\ \bibnamefont {Horowitz}},\ and\ \bibinfo
  {author} {\bibfnamefont {M.}~\bibnamefont {Vanderhaeghen}},\ }\bibfield
  {title} {\bibinfo {title} {Neutron skins of atomic nuclei: per aspera ad
  astra},\ }\href {https://doi.org/10.1088/1361-6471/ab2c6d} {\bibfield
  {journal} {\bibinfo  {journal} {J. Phys. G: Nucl. Part. Phys.}\ }\textbf
  {\bibinfo {volume} {46}},\ \bibinfo {pages} {093003} (\bibinfo {year}
  {2019})}\BibitemShut {NoStop}%
\bibitem [{\citenamefont {Brown}(2000)}]{bro00}%
  \BibitemOpen
  \bibfield  {author} {\bibinfo {author} {\bibfnamefont {B.~A.}\ \bibnamefont
  {Brown}},\ }\bibfield  {title} {\bibinfo {title} {Neutron radii in nuclei and
  the neutron equation of state},\ }\href
  {https://doi.org/10.1103/PhysRevLett.85.5296} {\bibfield  {journal} {\bibinfo
   {journal} {Phys. Rev. Lett.}\ }\textbf {\bibinfo {volume} {85}},\ \bibinfo
  {pages} {5296} (\bibinfo {year} {2000})}\BibitemShut {NoStop}%
\bibitem [{\citenamefont {Centelles}\ \emph {et~al.}(2009)\citenamefont
  {Centelles}, \citenamefont {Roca-Maza}, \citenamefont {Vi\~nas},\ and\
  \citenamefont {Warda}}]{cen09}%
  \BibitemOpen
  \bibfield  {author} {\bibinfo {author} {\bibfnamefont {M.}~\bibnamefont
  {Centelles}}, \bibinfo {author} {\bibfnamefont {X.}~\bibnamefont
  {Roca-Maza}}, \bibinfo {author} {\bibfnamefont {X.}~\bibnamefont {Vi\~nas}},\
  and\ \bibinfo {author} {\bibfnamefont {M.}~\bibnamefont {Warda}},\ }\bibfield
   {title} {\bibinfo {title} {Nuclear symmetry energy probed by neutron skin
  thickness of nuclei},\ }\href
  {https://doi.org/10.1103/PhysRevLett.102.122502} {\bibfield  {journal}
  {\bibinfo  {journal} {Phys. Rev. Lett.}\ }\textbf {\bibinfo {volume} {102}},\
  \bibinfo {pages} {122502} (\bibinfo {year} {2009})}\BibitemShut {NoStop}%
\bibitem [{\citenamefont {Reinhard}\ and\ \citenamefont
  {Nazarewicz}(2010)}]{rei10}%
  \BibitemOpen
  \bibfield  {author} {\bibinfo {author} {\bibfnamefont {P.-G.}\ \bibnamefont
  {Reinhard}}\ and\ \bibinfo {author} {\bibfnamefont {W.}~\bibnamefont
  {Nazarewicz}},\ }\bibfield  {title} {\bibinfo {title} {Information content of
  a new observable: {T}he case of the nuclear neutron skin},\ }\href
  {https://doi.org/10.1103/PhysRevC.81.051303} {\bibfield  {journal} {\bibinfo
  {journal} {Phys. Rev. C}\ }\textbf {\bibinfo {volume} {81}},\ \bibinfo
  {pages} {051303(R)} (\bibinfo {year} {2010})}\BibitemShut {NoStop}%
\bibitem [{\citenamefont {Roca-Maza}\ \emph {et~al.}(2011)\citenamefont
  {Roca-Maza}, \citenamefont {Centelles}, \citenamefont {Vi\~nas},\ and\
  \citenamefont {Warda}}]{roc11}%
  \BibitemOpen
  \bibfield  {author} {\bibinfo {author} {\bibfnamefont {X.}~\bibnamefont
  {Roca-Maza}}, \bibinfo {author} {\bibfnamefont {M.}~\bibnamefont
  {Centelles}}, \bibinfo {author} {\bibfnamefont {X.}~\bibnamefont {Vi\~nas}},\
  and\ \bibinfo {author} {\bibfnamefont {M.}~\bibnamefont {Warda}},\ }\bibfield
   {title} {\bibinfo {title} {Neutron skin of $^{208}\mathrm{Pb}$, nuclear
  symmetry energy, and the parity radius experiment},\ }\href
  {https://doi.org/10.1103/PhysRevLett.106.252501} {\bibfield  {journal}
  {\bibinfo  {journal} {Phys. Rev. Lett.}\ }\textbf {\bibinfo {volume} {106}},\
  \bibinfo {pages} {252501} (\bibinfo {year} {2011})}\BibitemShut {NoStop}%
\bibitem [{\citenamefont {Lattimer}(2023)}]{lat23}%
  \BibitemOpen
  \bibfield  {author} {\bibinfo {author} {\bibfnamefont {J.~M.}\ \bibnamefont
  {Lattimer}},\ }\bibfield  {title} {\bibinfo {title} {Constraints on nuclear
  symmetry energy parameters},\ }\href
  {https://doi.org/10.3390/particles6010003} {\bibfield  {journal} {\bibinfo
  {journal} {Particles}\ }\textbf {\bibinfo {volume} {6}},\ \bibinfo {pages}
  {30} (\bibinfo {year} {2023})}\BibitemShut {NoStop}%
\bibitem [{\citenamefont {Roca-Maza}\ \emph {et~al.}(2013)\citenamefont
  {Roca-Maza}, \citenamefont {Brenna}, \citenamefont {Col\`o}, \citenamefont
  {Centelles}, \citenamefont {Vi\~nas}, \citenamefont {Agrawal}, \citenamefont
  {Paar}, \citenamefont {Vretenar},\ and\ \citenamefont {Piekarewicz}}]{roc13}%
  \BibitemOpen
  \bibfield  {author} {\bibinfo {author} {\bibfnamefont {X.}~\bibnamefont
  {Roca-Maza}}, \bibinfo {author} {\bibfnamefont {M.}~\bibnamefont {Brenna}},
  \bibinfo {author} {\bibfnamefont {G.}~\bibnamefont {Col\`o}}, \bibinfo
  {author} {\bibfnamefont {M.}~\bibnamefont {Centelles}}, \bibinfo {author}
  {\bibfnamefont {X.}~\bibnamefont {Vi\~nas}}, \bibinfo {author} {\bibfnamefont
  {B.~K.}\ \bibnamefont {Agrawal}}, \bibinfo {author} {\bibfnamefont
  {N.}~\bibnamefont {Paar}}, \bibinfo {author} {\bibfnamefont {D.}~\bibnamefont
  {Vretenar}},\ and\ \bibinfo {author} {\bibfnamefont {J.}~\bibnamefont
  {Piekarewicz}},\ }\bibfield  {title} {\bibinfo {title} {Electric dipole
  polarizability in ${}^{208}${P}b: {I}nsights from the droplet model},\ }\href
  {https://doi.org/10.1103/PhysRevC.88.024316} {\bibfield  {journal} {\bibinfo
  {journal} {Phys. Rev. C}\ }\textbf {\bibinfo {volume} {88}},\ \bibinfo
  {pages} {024316} (\bibinfo {year} {2013})}\BibitemShut {NoStop}%
\bibitem [{\citenamefont {von Neumann-Cosel}\ and\ \citenamefont
  {Tamii}(2019)}]{vnc19a}%
  \BibitemOpen
  \bibfield  {author} {\bibinfo {author} {\bibfnamefont {P.}~\bibnamefont {von
  Neumann-Cosel}}\ and\ \bibinfo {author} {\bibfnamefont {A.}~\bibnamefont
  {Tamii}},\ }\bibfield  {title} {\bibinfo {title} {Electric and magnetic
  dipole modes in high-resolution inelastic proton scattering at
  0{\textdegree}},\ }\href {https://doi.org/10.1140/epja/i2019-12781-7}
  {\bibfield  {journal} {\bibinfo  {journal} {Eur. Phys. J. A}\ }\textbf
  {\bibinfo {volume} {55}},\ \bibinfo {pages} {110} (\bibinfo {year}
  {2019})}\BibitemShut {NoStop}%
\bibitem [{\citenamefont {Tamii}\ \emph {et~al.}(2011)\citenamefont {Tamii}
  \emph {et~al.}}]{tam11}%
  \BibitemOpen
  \bibfield  {author} {\bibinfo {author} {\bibfnamefont {A.}~\bibnamefont
  {Tamii}} \emph {et~al.},\ }\bibfield  {title} {\bibinfo {title} {Complete
  electric dipole response and the neutron skin in $^{208}\mathrm{Pb}$},\
  }\href {https://doi.org/10.1103/PhysRevLett.107.062502} {\bibfield  {journal}
  {\bibinfo  {journal} {Phys. Rev. Lett.}\ }\textbf {\bibinfo {volume} {107}},\
  \bibinfo {pages} {062502} (\bibinfo {year} {2011})}\BibitemShut {NoStop}%
\bibitem [{\citenamefont {Hashimoto}\ \emph {et~al.}(2015)\citenamefont
  {Hashimoto} \emph {et~al.}}]{has15}%
  \BibitemOpen
  \bibfield  {author} {\bibinfo {author} {\bibfnamefont {T.}~\bibnamefont
  {Hashimoto}} \emph {et~al.},\ }\bibfield  {title} {\bibinfo {title} {Dipole
  polarizability of $^{120}\mathrm{Sn}$ and nuclear energy density
  functionals},\ }\href {https://doi.org/10.1103/PhysRevC.92.031305} {\bibfield
   {journal} {\bibinfo  {journal} {Phys. Rev. C}\ }\textbf {\bibinfo {volume}
  {92}},\ \bibinfo {pages} {031305} (\bibinfo {year} {2015})}\BibitemShut
  {NoStop}%
\bibitem [{\citenamefont {Birkhan}\ \emph {et~al.}(2017)\citenamefont
  {Birkhan}, \citenamefont {Miorelli}, \citenamefont {Bacca}, \citenamefont
  {Bassauer}, \citenamefont {Bertulani}, \citenamefont {Hagen}, \citenamefont
  {Matsubara}, \citenamefont {von Neumann-Cosel}, \citenamefont {Papenbrock},
  \citenamefont {Pietralla}, \citenamefont {Ponomarev}, \citenamefont
  {Richter}, \citenamefont {Schwenk},\ and\ \citenamefont {Tamii}}]{bir17}%
  \BibitemOpen
  \bibfield  {author} {\bibinfo {author} {\bibfnamefont {J.}~\bibnamefont
  {Birkhan}}, \bibinfo {author} {\bibfnamefont {M.}~\bibnamefont {Miorelli}},
  \bibinfo {author} {\bibfnamefont {S.}~\bibnamefont {Bacca}}, \bibinfo
  {author} {\bibfnamefont {S.}~\bibnamefont {Bassauer}}, \bibinfo {author}
  {\bibfnamefont {C.~A.}\ \bibnamefont {Bertulani}}, \bibinfo {author}
  {\bibfnamefont {G.}~\bibnamefont {Hagen}}, \bibinfo {author} {\bibfnamefont
  {H.}~\bibnamefont {Matsubara}}, \bibinfo {author} {\bibfnamefont
  {P.}~\bibnamefont {von Neumann-Cosel}}, \bibinfo {author} {\bibfnamefont
  {T.}~\bibnamefont {Papenbrock}}, \bibinfo {author} {\bibfnamefont
  {N.}~\bibnamefont {Pietralla}}, \bibinfo {author} {\bibfnamefont {V.~Y.}\
  \bibnamefont {Ponomarev}}, \bibinfo {author} {\bibfnamefont {A.}~\bibnamefont
  {Richter}}, \bibinfo {author} {\bibfnamefont {A.}~\bibnamefont {Schwenk}},\
  and\ \bibinfo {author} {\bibfnamefont {A.}~\bibnamefont {Tamii}},\ }\bibfield
   {title} {\bibinfo {title} {Electric dipole polarizability of
  $^{48}\mathrm{Ca}$ and implications for the neutron skin},\ }\href
  {https://doi.org/10.1103/PhysRevLett.118.252501} {\bibfield  {journal}
  {\bibinfo  {journal} {Phys. Rev. Lett.}\ }\textbf {\bibinfo {volume} {118}},\
  \bibinfo {pages} {252501} (\bibinfo {year} {2017})}\BibitemShut {NoStop}%
\bibitem [{\citenamefont {Bassauer}\ \emph
  {et~al.}(2020{\natexlab{a}})\citenamefont {Bassauer} \emph
  {et~al.}}]{bas20a}%
  \BibitemOpen
  \bibfield  {author} {\bibinfo {author} {\bibfnamefont {S.}~\bibnamefont
  {Bassauer}} \emph {et~al.},\ }\bibfield  {title} {\bibinfo {title} {Evolution
  of the dipole polarizability in the stable tin isotope chain},\ }\href
  {https://doi.org/https://doi.org/10.1016/j.physletb.2020.135804} {\bibfield
  {journal} {\bibinfo  {journal} {Phys. Lett. B}\ }\textbf {\bibinfo {volume}
  {810}},\ \bibinfo {pages} {135804} (\bibinfo {year}
  {2020}{\natexlab{a}})}\BibitemShut {NoStop}%
\bibitem [{\citenamefont {Fearick}\ \emph {et~al.}(2023)\citenamefont
  {Fearick}, \citenamefont {von Neumann-Cosel}, \citenamefont {Bacca},
  \citenamefont {Birkhan}, \citenamefont {Bonaiti}, \citenamefont {Brandherm},
  \citenamefont {Hagen}, \citenamefont {Matsubara}, \citenamefont {Nazarewicz},
  \citenamefont {Pietralla}, \citenamefont {Ponomarev}, \citenamefont
  {Reinhard}, \citenamefont {Roca-Maza}, \citenamefont {Richter}, \citenamefont
  {Schwenk}, \citenamefont {Simonis},\ and\ \citenamefont {Tamii}}]{fea23}%
  \BibitemOpen
  \bibfield  {author} {\bibinfo {author} {\bibfnamefont {R.~W.}\ \bibnamefont
  {Fearick}}, \bibinfo {author} {\bibfnamefont {P.}~\bibnamefont {von
  Neumann-Cosel}}, \bibinfo {author} {\bibfnamefont {S.}~\bibnamefont {Bacca}},
  \bibinfo {author} {\bibfnamefont {J.}~\bibnamefont {Birkhan}}, \bibinfo
  {author} {\bibfnamefont {F.}~\bibnamefont {Bonaiti}}, \bibinfo {author}
  {\bibfnamefont {I.}~\bibnamefont {Brandherm}}, \bibinfo {author}
  {\bibfnamefont {G.}~\bibnamefont {Hagen}}, \bibinfo {author} {\bibfnamefont
  {H.}~\bibnamefont {Matsubara}}, \bibinfo {author} {\bibfnamefont
  {W.}~\bibnamefont {Nazarewicz}}, \bibinfo {author} {\bibfnamefont
  {N.}~\bibnamefont {Pietralla}}, \bibinfo {author} {\bibfnamefont {V.~Y.}\
  \bibnamefont {Ponomarev}}, \bibinfo {author} {\bibfnamefont {P.-G.}\
  \bibnamefont {Reinhard}}, \bibinfo {author} {\bibfnamefont {X.}~\bibnamefont
  {Roca-Maza}}, \bibinfo {author} {\bibfnamefont {A.}~\bibnamefont {Richter}},
  \bibinfo {author} {\bibfnamefont {A.}~\bibnamefont {Schwenk}}, \bibinfo
  {author} {\bibfnamefont {J.}~\bibnamefont {Simonis}},\ and\ \bibinfo {author}
  {\bibfnamefont {A.}~\bibnamefont {Tamii}},\ }\bibfield  {title} {\bibinfo
  {title} {Electric dipole polarizability of $^{40}\mathrm{Ca}$},\ }\href
  {https://doi.org/10.1103/PhysRevResearch.5.L022044} {\bibfield  {journal}
  {\bibinfo  {journal} {Phys. Rev. Res.}\ }\textbf {\bibinfo {volume} {5}},\
  \bibinfo {pages} {L022044} (\bibinfo {year} {2023})}\BibitemShut {NoStop}%
\bibitem [{\citenamefont {Bender}\ \emph {et~al.}(2003)\citenamefont {Bender},
  \citenamefont {Heenen},\ and\ \citenamefont {Reinhard}}]{ben03}%
  \BibitemOpen
  \bibfield  {author} {\bibinfo {author} {\bibfnamefont {M.}~\bibnamefont
  {Bender}}, \bibinfo {author} {\bibfnamefont {P.-H.}\ \bibnamefont {Heenen}},\
  and\ \bibinfo {author} {\bibfnamefont {P.-G.}\ \bibnamefont {Reinhard}},\
  }\bibfield  {title} {\bibinfo {title} {Self-consistent mean-field models for
  nuclear structure},\ }\href {https://doi.org/10.1103/RevModPhys.75.121}
  {\bibfield  {journal} {\bibinfo  {journal} {Rev. Mod. Phys.}\ }\textbf
  {\bibinfo {volume} {75}},\ \bibinfo {pages} {121} (\bibinfo {year}
  {2003})}\BibitemShut {NoStop}%
\bibitem [{\citenamefont {Piekarewicz}\ \emph {et~al.}(2012)\citenamefont
  {Piekarewicz}, \citenamefont {Agrawal}, \citenamefont {Col\`o}, \citenamefont
  {Nazarewicz}, \citenamefont {Paar}, \citenamefont {Reinhard}, \citenamefont
  {Roca-Maza},\ and\ \citenamefont {Vretenar}}]{piekarewicz2012}%
  \BibitemOpen
  \bibfield  {author} {\bibinfo {author} {\bibfnamefont {J.}~\bibnamefont
  {Piekarewicz}}, \bibinfo {author} {\bibfnamefont {B.~K.}\ \bibnamefont
  {Agrawal}}, \bibinfo {author} {\bibfnamefont {G.}~\bibnamefont {Col\`o}},
  \bibinfo {author} {\bibfnamefont {W.}~\bibnamefont {Nazarewicz}}, \bibinfo
  {author} {\bibfnamefont {N.}~\bibnamefont {Paar}}, \bibinfo {author}
  {\bibfnamefont {P.-G.}\ \bibnamefont {Reinhard}}, \bibinfo {author}
  {\bibfnamefont {X.}~\bibnamefont {Roca-Maza}},\ and\ \bibinfo {author}
  {\bibfnamefont {D.}~\bibnamefont {Vretenar}},\ }\bibfield  {title} {\bibinfo
  {title} {Electric dipole polarizability and the neutron skin},\ }\href
  {https://doi.org/10.1103/PhysRevC.85.041302} {\bibfield  {journal} {\bibinfo
  {journal} {Phys. Rev. C}\ }\textbf {\bibinfo {volume} {85}},\ \bibinfo
  {pages} {041302} (\bibinfo {year} {2012})}\BibitemShut {NoStop}%
\bibitem [{\citenamefont {Hagen}\ \emph {et~al.}(2016)\citenamefont {Hagen},
  \citenamefont {Ekstr{\"o}m}, \citenamefont {Forss{\'e}n}, \citenamefont
  {Jansen}, \citenamefont {Nazarewicz}, \citenamefont {Papenbrock},
  \citenamefont {Wendt}, \citenamefont {Bacca}, \citenamefont {Barnea},
  \citenamefont {Carlsson}, \citenamefont {Drischler}, \citenamefont {Hebeler},
  \citenamefont {Hjorth-Jensen}, \citenamefont {Miorelli}, \citenamefont
  {Orlandini}, \citenamefont {Schwenk},\ and\ \citenamefont {Simonis}}]{hag16}%
  \BibitemOpen
  \bibfield  {author} {\bibinfo {author} {\bibfnamefont {G.}~\bibnamefont
  {Hagen}}, \bibinfo {author} {\bibfnamefont {A.}~\bibnamefont {Ekstr{\"o}m}},
  \bibinfo {author} {\bibfnamefont {C.}~\bibnamefont {Forss{\'e}n}}, \bibinfo
  {author} {\bibfnamefont {G.~R.}\ \bibnamefont {Jansen}}, \bibinfo {author}
  {\bibfnamefont {W.}~\bibnamefont {Nazarewicz}}, \bibinfo {author}
  {\bibfnamefont {T.}~\bibnamefont {Papenbrock}}, \bibinfo {author}
  {\bibfnamefont {K.~A.}\ \bibnamefont {Wendt}}, \bibinfo {author}
  {\bibfnamefont {S.}~\bibnamefont {Bacca}}, \bibinfo {author} {\bibfnamefont
  {N.}~\bibnamefont {Barnea}}, \bibinfo {author} {\bibfnamefont
  {B.}~\bibnamefont {Carlsson}}, \bibinfo {author} {\bibfnamefont
  {C.}~\bibnamefont {Drischler}}, \bibinfo {author} {\bibfnamefont
  {K.}~\bibnamefont {Hebeler}}, \bibinfo {author} {\bibfnamefont
  {M.}~\bibnamefont {Hjorth-Jensen}}, \bibinfo {author} {\bibfnamefont
  {M.}~\bibnamefont {Miorelli}}, \bibinfo {author} {\bibfnamefont
  {G.}~\bibnamefont {Orlandini}}, \bibinfo {author} {\bibfnamefont
  {A.}~\bibnamefont {Schwenk}},\ and\ \bibinfo {author} {\bibfnamefont
  {J.}~\bibnamefont {Simonis}},\ }\bibfield  {title} {\bibinfo {title} {Neutron
  and weak-charge distributions of the $^{48}${C}a nucleus},\ }\href
  {https://doi.org/10.1038/nphys3529} {\bibfield  {journal} {\bibinfo
  {journal} {Nat. Phys.}\ }\textbf {\bibinfo {volume} {12}},\ \bibinfo {pages}
  {186} (\bibinfo {year} {2016})}\BibitemShut {NoStop}%
\bibitem [{\citenamefont {Simonis}\ \emph {et~al.}(2019)\citenamefont
  {Simonis}, \citenamefont {Bacca},\ and\ \citenamefont {Hagen}}]{sim19}%
  \BibitemOpen
  \bibfield  {author} {\bibinfo {author} {\bibfnamefont {J.}~\bibnamefont
  {Simonis}}, \bibinfo {author} {\bibfnamefont {S.}~\bibnamefont {Bacca}},\
  and\ \bibinfo {author} {\bibfnamefont {G.}~\bibnamefont {Hagen}},\ }\bibfield
   {title} {\bibinfo {title} {First principles electromagnetic responses in
  medium-mass nuclei},\ }\href@noop {} {\bibfield  {journal} {\bibinfo
  {journal} {Eur. Phys. J. A}\ }\textbf {\bibinfo {volume} {55}},\ \bibinfo
  {pages} {241} (\bibinfo {year} {2019})}\BibitemShut {NoStop}%
\bibitem [{\citenamefont {Hu}\ \emph {et~al.}(2022)\citenamefont {Hu},
  \citenamefont {Jiang}, \citenamefont {Miyagi}, \citenamefont {Sun},
  \citenamefont {Ekstr{\"o}m}, \citenamefont {Forss{\'e}n}, \citenamefont
  {Hagen}, \citenamefont {Holt}, \citenamefont {Papenbrock}, \citenamefont
  {Stroberg},\ and\ \citenamefont {Vernon}}]{hu21}%
  \BibitemOpen
  \bibfield  {author} {\bibinfo {author} {\bibfnamefont {B.}~\bibnamefont
  {Hu}}, \bibinfo {author} {\bibfnamefont {W.}~\bibnamefont {Jiang}}, \bibinfo
  {author} {\bibfnamefont {T.}~\bibnamefont {Miyagi}}, \bibinfo {author}
  {\bibfnamefont {Z.}~\bibnamefont {Sun}}, \bibinfo {author} {\bibfnamefont
  {A.}~\bibnamefont {Ekstr{\"o}m}}, \bibinfo {author} {\bibfnamefont
  {C.}~\bibnamefont {Forss{\'e}n}}, \bibinfo {author} {\bibfnamefont
  {G.}~\bibnamefont {Hagen}}, \bibinfo {author} {\bibfnamefont {J.~D.}\
  \bibnamefont {Holt}}, \bibinfo {author} {\bibfnamefont {T.}~\bibnamefont
  {Papenbrock}}, \bibinfo {author} {\bibfnamefont {S.~R.}\ \bibnamefont
  {Stroberg}},\ and\ \bibinfo {author} {\bibfnamefont {I.}~\bibnamefont
  {Vernon}},\ }\bibfield  {title} {\bibinfo {title} {Ab initio predictions link
  the neutron skin of $^{208}${P}b to nuclear forces},\ }\href
  {https://doi.org/10.1038/s41567-022-01715-8} {\bibfield  {journal} {\bibinfo
  {journal} {Nat. Phys.}\ }\textbf {\bibinfo {volume} {18}},\ \bibinfo {pages}
  {1196} (\bibinfo {year} {2022})}\BibitemShut {NoStop}%
\bibitem [{\citenamefont {Hebeler}\ \emph {et~al.}(2011)\citenamefont
  {Hebeler}, \citenamefont {Bogner}, \citenamefont {Furnstahl}, \citenamefont
  {Nogga},\ and\ \citenamefont {Schwenk}}]{heb11}%
  \BibitemOpen
  \bibfield  {author} {\bibinfo {author} {\bibfnamefont {K.}~\bibnamefont
  {Hebeler}}, \bibinfo {author} {\bibfnamefont {S.~K.}\ \bibnamefont {Bogner}},
  \bibinfo {author} {\bibfnamefont {R.~J.}\ \bibnamefont {Furnstahl}}, \bibinfo
  {author} {\bibfnamefont {A.}~\bibnamefont {Nogga}},\ and\ \bibinfo {author}
  {\bibfnamefont {A.}~\bibnamefont {Schwenk}},\ }\bibfield  {title} {\bibinfo
  {title} {Improved nuclear matter calculations from chiral low-momentum
  interactions},\ }\href {https://doi.org/10.1103/PhysRevC.83.031301}
  {\bibfield  {journal} {\bibinfo  {journal} {Phys. Rev. C}\ }\textbf {\bibinfo
  {volume} {83}},\ \bibinfo {pages} {031301(R)} (\bibinfo {year}
  {2011})}\BibitemShut {NoStop}%
\bibitem [{\citenamefont {Ekstr\"om}\ \emph {et~al.}(2015)\citenamefont
  {Ekstr\"om}, \citenamefont {Jansen}, \citenamefont {Wendt}, \citenamefont
  {Hagen}, \citenamefont {Papenbrock}, \citenamefont {Carlsson}, \citenamefont
  {Forss\'en}, \citenamefont {Hjorth-Jensen}, \citenamefont {Navr\'atil},\ and\
  \citenamefont {Nazarewicz}}]{eks15}%
  \BibitemOpen
  \bibfield  {author} {\bibinfo {author} {\bibfnamefont {A.}~\bibnamefont
  {Ekstr\"om}}, \bibinfo {author} {\bibfnamefont {G.~R.}\ \bibnamefont
  {Jansen}}, \bibinfo {author} {\bibfnamefont {K.~A.}\ \bibnamefont {Wendt}},
  \bibinfo {author} {\bibfnamefont {G.}~\bibnamefont {Hagen}}, \bibinfo
  {author} {\bibfnamefont {T.}~\bibnamefont {Papenbrock}}, \bibinfo {author}
  {\bibfnamefont {B.~D.}\ \bibnamefont {Carlsson}}, \bibinfo {author}
  {\bibfnamefont {C.}~\bibnamefont {Forss\'en}}, \bibinfo {author}
  {\bibfnamefont {M.}~\bibnamefont {Hjorth-Jensen}}, \bibinfo {author}
  {\bibfnamefont {P.}~\bibnamefont {Navr\'atil}},\ and\ \bibinfo {author}
  {\bibfnamefont {W.}~\bibnamefont {Nazarewicz}},\ }\bibfield  {title}
  {\bibinfo {title} {Accurate nuclear radii and binding energies from a chiral
  interaction},\ }\href {https://doi.org/10.1103/PhysRevC.91.051301} {\bibfield
   {journal} {\bibinfo  {journal} {Phys. Rev. C}\ }\textbf {\bibinfo {volume}
  {91}},\ \bibinfo {pages} {051301(R)} (\bibinfo {year} {2015})}\BibitemShut
  {NoStop}%
\bibitem [{\citenamefont {Roca-Maza}\ \emph {et~al.}(2015)\citenamefont
  {Roca-Maza}, \citenamefont {Vi\~nas}, \citenamefont {Centelles},
  \citenamefont {Agrawal}, \citenamefont {Col\`o}, \citenamefont {Paar},
  \citenamefont {Piekarewicz},\ and\ \citenamefont {Vretenar}}]{roc15}%
  \BibitemOpen
  \bibfield  {author} {\bibinfo {author} {\bibfnamefont {X.}~\bibnamefont
  {Roca-Maza}}, \bibinfo {author} {\bibfnamefont {X.}~\bibnamefont {Vi\~nas}},
  \bibinfo {author} {\bibfnamefont {M.}~\bibnamefont {Centelles}}, \bibinfo
  {author} {\bibfnamefont {B.~K.}\ \bibnamefont {Agrawal}}, \bibinfo {author}
  {\bibfnamefont {G.}~\bibnamefont {Col\`o}}, \bibinfo {author} {\bibfnamefont
  {N.}~\bibnamefont {Paar}}, \bibinfo {author} {\bibfnamefont {J.}~\bibnamefont
  {Piekarewicz}},\ and\ \bibinfo {author} {\bibfnamefont {D.}~\bibnamefont
  {Vretenar}},\ }\bibfield  {title} {\bibinfo {title} {Neutron skin thickness
  from the measured electric dipole polarizability in $^{68}\text{Ni}$,
  $^{120}\text{Sn}$, and $^{208}\text{Pb}$},\ }\href
  {https://doi.org/10.1103/PhysRevC.92.064304} {\bibfield  {journal} {\bibinfo
  {journal} {Phys. Rev. C}\ }\textbf {\bibinfo {volume} {92}},\ \bibinfo
  {pages} {064304} (\bibinfo {year} {2015})}\BibitemShut {NoStop}%
\bibitem [{\citenamefont {Kaufmann}\ \emph {et~al.}(2020)\citenamefont
  {Kaufmann} \emph {et~al.}}]{kau20}%
  \BibitemOpen
  \bibfield  {author} {\bibinfo {author} {\bibfnamefont {S.}~\bibnamefont
  {Kaufmann}} \emph {et~al.},\ }\bibfield  {title} {\bibinfo {title} {Charge
  radius of the short-lived $^{68}\mathrm{Ni}$ and correlation with the dipole
  polarizability},\ }\href {https://doi.org/10.1103/PhysRevLett.124.132502}
  {\bibfield  {journal} {\bibinfo  {journal} {Phys. Rev. Lett.}\ }\textbf
  {\bibinfo {volume} {124}},\ \bibinfo {pages} {132502} (\bibinfo {year}
  {2020})}\BibitemShut {NoStop}%
\bibitem [{\citenamefont {Bonaiti}\ \emph {et~al.}(2022)\citenamefont
  {Bonaiti}, \citenamefont {Bacca},\ and\ \citenamefont {Hagen}}]{bonaiti2022}%
  \BibitemOpen
  \bibfield  {author} {\bibinfo {author} {\bibfnamefont {F.}~\bibnamefont
  {Bonaiti}}, \bibinfo {author} {\bibfnamefont {S.}~\bibnamefont {Bacca}},\
  and\ \bibinfo {author} {\bibfnamefont {G.}~\bibnamefont {Hagen}},\ }\bibfield
   {title} {\bibinfo {title} {Ab initio coupled-cluster calculations of ground
  and dipole excited states in $^{8}\mathrm{He}$},\ }\href
  {https://doi.org/10.1103/PhysRevC.105.034313} {\bibfield  {journal} {\bibinfo
   {journal} {Phys. Rev. C}\ }\textbf {\bibinfo {volume} {105}},\ \bibinfo
  {pages} {034313} (\bibinfo {year} {2022})}\BibitemShut {NoStop}%
\bibitem [{\citenamefont {Bonaiti}\ and\ \citenamefont
  {Bacca}(2024)}]{bonaiti2024_fbs}%
  \BibitemOpen
  \bibfield  {author} {\bibinfo {author} {\bibfnamefont {F.}~\bibnamefont
  {Bonaiti}}\ and\ \bibinfo {author} {\bibfnamefont {S.}~\bibnamefont
  {Bacca}},\ }\bibfield  {title} {\bibinfo {title} {Low-energy dipole strength
  in $^8${H}e},\ }\href {https://doi.org/10.1007/s00601-024-01903-7} {\bibfield
   {journal} {\bibinfo  {journal} {Few-Body Syst.}\ }\textbf {\bibinfo {volume}
  {65}},\ \bibinfo {pages} {54} (\bibinfo {year} {2024})}\BibitemShut {NoStop}%
\bibitem [{\citenamefont {Bonaiti}\ \emph {et~al.}(2024)\citenamefont
  {Bonaiti}, \citenamefont {Bacca}, \citenamefont {Hagen},\ and\ \citenamefont
  {Jansen}}]{bonaiti2024}%
  \BibitemOpen
  \bibfield  {author} {\bibinfo {author} {\bibfnamefont {F.}~\bibnamefont
  {Bonaiti}}, \bibinfo {author} {\bibfnamefont {S.}~\bibnamefont {Bacca}},
  \bibinfo {author} {\bibfnamefont {G.}~\bibnamefont {Hagen}},\ and\ \bibinfo
  {author} {\bibfnamefont {G.~R.}\ \bibnamefont {Jansen}},\ }\bibfield  {title}
  {\bibinfo {title} {{Electromagnetic observables of open-shell nuclei from
  coupled-cluster theory}},\ }\href@noop {} {\  (\bibinfo {year} {2024})},\
  \Eprint {https://arxiv.org/abs/2405.05608} {arXiv:2405.05608 [nucl-th]}
  \BibitemShut {NoStop}%
\bibitem [{\citenamefont {Rossi}\ \emph {et~al.}(2013)\citenamefont {Rossi}
  \emph {et~al.}}]{ros13}%
  \BibitemOpen
  \bibfield  {author} {\bibinfo {author} {\bibfnamefont {D.~M.}\ \bibnamefont
  {Rossi}} \emph {et~al.},\ }\bibfield  {title} {\bibinfo {title} {Measurement
  of the dipole polarizability of the unstable neutron-rich nucleus
  $^{68}\mathrm{Ni}$},\ }\href {https://doi.org/10.1103/PhysRevLett.111.242503}
  {\bibfield  {journal} {\bibinfo  {journal} {Phys. Rev. Lett.}\ }\textbf
  {\bibinfo {volume} {111}},\ \bibinfo {pages} {242503} (\bibinfo {year}
  {2013})}\BibitemShut {NoStop}%
\bibitem [{\citenamefont {Tamii}\ \emph {et~al.}(2009)\citenamefont {Tamii}
  \emph {et~al.}}]{tam09}%
  \BibitemOpen
  \bibfield  {author} {\bibinfo {author} {\bibfnamefont {A.}~\bibnamefont
  {Tamii}} \emph {et~al.},\ }\bibfield  {title} {\bibinfo {title} {Measurement
  of high energy resolution inelastic proton scattering at and close to zero
  degrees},\ }\href {https://doi.org/doi:10.1016/j.nima.2009.03.248} {\bibfield
   {journal} {\bibinfo  {journal} {Nucl. Instrum. Methods Phys. Res., Sect. A}\
  }\textbf {\bibinfo {volume} {605}},\ \bibinfo {pages} {326 } (\bibinfo {year}
  {2009})}\BibitemShut {NoStop}%
\bibitem [{\citenamefont {Brandherm}\ \emph {et~al.}(2024)\citenamefont
  {Brandherm}, \citenamefont {von Neumann-Cosel}, \citenamefont {Mancino},
  \citenamefont {Mart\'{\i}nez-Pinedo}, \citenamefont {Matsubara},
  \citenamefont {Ponomarev}, \citenamefont {Richter}, \citenamefont {Scheck},\
  and\ \citenamefont {Tamii}}]{bra24}%
  \BibitemOpen
  \bibfield  {author} {\bibinfo {author} {\bibfnamefont {I.}~\bibnamefont
  {Brandherm}}, \bibinfo {author} {\bibfnamefont {P.}~\bibnamefont {von
  Neumann-Cosel}}, \bibinfo {author} {\bibfnamefont {R.}~\bibnamefont
  {Mancino}}, \bibinfo {author} {\bibfnamefont {G.}~\bibnamefont
  {Mart\'{\i}nez-Pinedo}}, \bibinfo {author} {\bibfnamefont {H.}~\bibnamefont
  {Matsubara}}, \bibinfo {author} {\bibfnamefont {V.~Y.}\ \bibnamefont
  {Ponomarev}}, \bibinfo {author} {\bibfnamefont {A.}~\bibnamefont {Richter}},
  \bibinfo {author} {\bibfnamefont {M.}~\bibnamefont {Scheck}},\ and\ \bibinfo
  {author} {\bibfnamefont {A.}~\bibnamefont {Tamii}},\ }\bibfield  {title}
  {\bibinfo {title} {Electric and magnetic dipole strength in
  $^{58}\mathrm{Ni}$ from forward-angle proton scattering},\ }\href
  {https://doi.org/10.1103/PhysRevC.110.034319} {\bibfield  {journal} {\bibinfo
   {journal} {Phys. Rev. C}\ }\textbf {\bibinfo {volume} {110}},\ \bibinfo
  {pages} {034319} (\bibinfo {year} {2024})}\BibitemShut {NoStop}%
\bibitem [{\citenamefont {von Neumann-Cosel}\ \emph {et~al.}(2023)\citenamefont
  {von Neumann-Cosel}, \citenamefont {Nesterenko}, \citenamefont {Brandherm},
  \citenamefont {Vishnevskiy}, \citenamefont {Reinhard}, \citenamefont
  {Kvasil}, \citenamefont {Matsubara}, \citenamefont {Repko}, \citenamefont
  {Richter}, \citenamefont {Scheck},\ and\ \citenamefont {Tamii}}]{vnc23}%
  \BibitemOpen
  \bibfield  {author} {\bibinfo {author} {\bibfnamefont {P.}~\bibnamefont {von
  Neumann-Cosel}}, \bibinfo {author} {\bibfnamefont {V.~O.}\ \bibnamefont
  {Nesterenko}}, \bibinfo {author} {\bibfnamefont {I.}~\bibnamefont
  {Brandherm}}, \bibinfo {author} {\bibfnamefont {P.~I.}\ \bibnamefont
  {Vishnevskiy}}, \bibinfo {author} {\bibfnamefont {P.-G.}\ \bibnamefont
  {Reinhard}}, \bibinfo {author} {\bibfnamefont {J.}~\bibnamefont {Kvasil}},
  \bibinfo {author} {\bibfnamefont {H.}~\bibnamefont {Matsubara}}, \bibinfo
  {author} {\bibfnamefont {A.}~\bibnamefont {Repko}}, \bibinfo {author}
  {\bibfnamefont {A.}~\bibnamefont {Richter}}, \bibinfo {author} {\bibfnamefont
  {M.}~\bibnamefont {Scheck}},\ and\ \bibinfo {author} {\bibfnamefont
  {A.}~\bibnamefont {Tamii}},\ }\bibfield  {title} {\bibinfo {title} {{Evidence
  for a Toroidal Electric Dipole Mode in Nuclei}},\ }\href@noop {} {\
  (\bibinfo {year} {2023})},\ \Eprint {https://arxiv.org/abs/2310.04736}
  {arXiv:2310.04736 [nucl-ex]} \BibitemShut {NoStop}%
\bibitem [{\citenamefont {Bassauer}\ \emph
  {et~al.}(2020{\natexlab{b}})\citenamefont {Bassauer} \emph
  {et~al.}}]{bas20b}%
  \BibitemOpen
  \bibfield  {author} {\bibinfo {author} {\bibfnamefont {S.}~\bibnamefont
  {Bassauer}} \emph {et~al.},\ }\bibfield  {title} {\bibinfo {title} {Electric
  and magnetic dipole strength in $^{112,114,116,118,120,124}\mathrm{Sn}$},\
  }\href {https://doi.org/10.1103/PhysRevC.102.034327} {\bibfield  {journal}
  {\bibinfo  {journal} {Phys. Rev. C}\ }\textbf {\bibinfo {volume} {102}},\
  \bibinfo {pages} {034327} (\bibinfo {year} {2020}{\natexlab{b}})}\BibitemShut
  {NoStop}%
\bibitem [{\citenamefont {Donaldson}\ \emph {et~al.}(2018)\citenamefont
  {Donaldson} \emph {et~al.}}]{don18}%
  \BibitemOpen
  \bibfield  {author} {\bibinfo {author} {\bibfnamefont {L.~M.}\ \bibnamefont
  {Donaldson}} \emph {et~al.},\ }\bibfield  {title} {\bibinfo {title}
  {Deformation dependence of the isovector giant dipole resonance: The
  neodymium isotopic chain revisited},\ }\href
  {https://doi.org/https://doi.org/10.1016/j.physletb.2017.11.025} {\bibfield
  {journal} {\bibinfo  {journal} {Phys. Lett. B}\ }\textbf {\bibinfo {volume}
  {776}},\ \bibinfo {pages} {133} (\bibinfo {year} {2018})}\BibitemShut
  {NoStop}%
\bibitem [{\citenamefont {Lui}\ \emph {et~al.}(2006)\citenamefont {Lui},
  \citenamefont {Youngblood}, \citenamefont {Clark}, \citenamefont {Tokimoto},\
  and\ \citenamefont {John}}]{liu06}%
  \BibitemOpen
  \bibfield  {author} {\bibinfo {author} {\bibfnamefont {Y.-W.}\ \bibnamefont
  {Lui}}, \bibinfo {author} {\bibfnamefont {D.~H.}\ \bibnamefont {Youngblood}},
  \bibinfo {author} {\bibfnamefont {H.~L.}\ \bibnamefont {Clark}}, \bibinfo
  {author} {\bibfnamefont {Y.}~\bibnamefont {Tokimoto}},\ and\ \bibinfo
  {author} {\bibfnamefont {B.}~\bibnamefont {John}},\ }\bibfield  {title}
  {\bibinfo {title} {Isoscalar giant resonances for nuclei with mass between 56
  and 60},\ }\href {https://doi.org/10.1103/PhysRevC.73.014314} {\bibfield
  {journal} {\bibinfo  {journal} {Phys. Rev. C}\ }\textbf {\bibinfo {volume}
  {73}},\ \bibinfo {pages} {014314} (\bibinfo {year} {2006})}\BibitemShut
  {NoStop}%
\bibitem [{\citenamefont {Li}\ \emph {et~al.}(2023)\citenamefont {Li},
  \citenamefont {Niu},\ and\ \citenamefont {Col\`o}}]{li23}%
  \BibitemOpen
  \bibfield  {author} {\bibinfo {author} {\bibfnamefont {Z.~Z.}\ \bibnamefont
  {Li}}, \bibinfo {author} {\bibfnamefont {Y.~F.}\ \bibnamefont {Niu}},\ and\
  \bibinfo {author} {\bibfnamefont {G.}~\bibnamefont {Col\`o}},\ }\bibfield
  {title} {\bibinfo {title} {Toward a unified description of isoscalar giant
  monopole resonances in a self-consistent quasiparticle-vibration coupling
  approach},\ }\href {https://doi.org/10.1103/PhysRevLett.131.082501}
  {\bibfield  {journal} {\bibinfo  {journal} {Phys. Rev. Lett.}\ }\textbf
  {\bibinfo {volume} {131}},\ \bibinfo {pages} {082501} (\bibinfo {year}
  {2023})}\BibitemShut {NoStop}%
\bibitem [{\citenamefont {Agrawal}\ \emph {et~al.}(2005)\citenamefont
  {Agrawal}, \citenamefont {Shlomo},\ and\ \citenamefont {Au}}]{agr05}%
  \BibitemOpen
  \bibfield  {author} {\bibinfo {author} {\bibfnamefont {B.~K.}\ \bibnamefont
  {Agrawal}}, \bibinfo {author} {\bibfnamefont {S.}~\bibnamefont {Shlomo}},\
  and\ \bibinfo {author} {\bibfnamefont {V.~K.}\ \bibnamefont {Au}},\
  }\bibfield  {title} {\bibinfo {title} {Determination of the parameters of a
  {S}kyrme type effective interaction using the simulated annealing approach},\
  }\href {https://doi.org/10.1103/PhysRevC.72.014310} {\bibfield  {journal}
  {\bibinfo  {journal} {Phys. Rev. C}\ }\textbf {\bibinfo {volume} {72}},\
  \bibinfo {pages} {014310} (\bibinfo {year} {2005})}\BibitemShut {NoStop}%
\bibitem [{\citenamefont {Bertulani}\ and\ \citenamefont {Baur}(1988)}]{ber88}%
  \BibitemOpen
  \bibfield  {author} {\bibinfo {author} {\bibfnamefont {C.~A.}\ \bibnamefont
  {Bertulani}}\ and\ \bibinfo {author} {\bibfnamefont {G.}~\bibnamefont
  {Baur}},\ }\bibfield  {title} {\bibinfo {title} {Electromagnetic processes in
  relativistic heavy ion collisions},\ }\href
  {https://doi.org/doi:10.1016/0370-1573(88)90142-1} {\bibfield  {journal}
  {\bibinfo  {journal} {Phys. Rep.}\ }\textbf {\bibinfo {volume} {163}},\
  \bibinfo {pages} {299} (\bibinfo {year} {1988})}\BibitemShut {NoStop}%
\bibitem [{\citenamefont {Bertulani}\ and\ \citenamefont
  {Nathan}(1993)}]{ber93}%
  \BibitemOpen
  \bibfield  {author} {\bibinfo {author} {\bibfnamefont {C.~A.}\ \bibnamefont
  {Bertulani}}\ and\ \bibinfo {author} {\bibfnamefont {A.~M.}\ \bibnamefont
  {Nathan}},\ }\bibfield  {title} {\bibinfo {title} {Excitation and photon
  decay of giant resonances from high-energy collisions of heavy ions},\ }\href
  {https://doi.org/doi:10.1016/0375-9474(93)90363-3} {\bibfield  {journal}
  {\bibinfo  {journal} {Nucl. Phys. A}\ }\textbf {\bibinfo {volume} {554}},\
  \bibinfo {pages} {158} (\bibinfo {year} {1993})}\BibitemShut {NoStop}%
\bibitem [{\citenamefont {Fultz}\ \emph {et~al.}(1974)\citenamefont {Fultz},
  \citenamefont {Alvarez}, \citenamefont {Berman},\ and\ \citenamefont
  {Meyer}}]{ful74}%
  \BibitemOpen
  \bibfield  {author} {\bibinfo {author} {\bibfnamefont {S.~C.}\ \bibnamefont
  {Fultz}}, \bibinfo {author} {\bibfnamefont {R.~A.}\ \bibnamefont {Alvarez}},
  \bibinfo {author} {\bibfnamefont {B.~L.}\ \bibnamefont {Berman}},\ and\
  \bibinfo {author} {\bibfnamefont {P.}~\bibnamefont {Meyer}},\ }\bibfield
  {title} {\bibinfo {title} {Photoneutron cross sections of $^{58}\mathrm{Ni}$
  and $^{60}\mathrm{Ni}$},\ }\href {https://doi.org/10.1103/PhysRevC.10.608}
  {\bibfield  {journal} {\bibinfo  {journal} {Phys. Rev. C}\ }\textbf {\bibinfo
  {volume} {10}},\ \bibinfo {pages} {608} (\bibinfo {year} {1974})}\BibitemShut
  {NoStop}%
\bibitem [{\citenamefont {Litvinova}(2023)}]{lit23}%
  \BibitemOpen
  \bibfield  {author} {\bibinfo {author} {\bibfnamefont {E.}~\bibnamefont
  {Litvinova}},\ }\bibfield  {title} {\bibinfo {title} {Relativistic approach
  to the nuclear breathing mode},\ }\href
  {https://doi.org/10.1103/PhysRevC.107.L041302} {\bibfield  {journal}
  {\bibinfo  {journal} {Phys. Rev. C}\ }\textbf {\bibinfo {volume} {107}},\
  \bibinfo {pages} {L041302} (\bibinfo {year} {2023})}\BibitemShut {NoStop}%
\bibitem [{\citenamefont {Garg}\ and\ \citenamefont {Col\`o}(2018)}]{gar18}%
  \BibitemOpen
  \bibfield  {author} {\bibinfo {author} {\bibfnamefont {U.}~\bibnamefont
  {Garg}}\ and\ \bibinfo {author} {\bibfnamefont {G.}~\bibnamefont {Col\`o}},\
  }\bibfield  {title} {\bibinfo {title} {The compression-mode giant resonances
  and nuclear incompressibility},\ }\href
  {https://doi.org/https://doi.org/10.1016/j.ppnp.2018.03.001} {\bibfield
  {journal} {\bibinfo  {journal} {Prog. Part. Nucl. Phys.}\ }\textbf {\bibinfo
  {volume} {101}},\ \bibinfo {pages} {55} (\bibinfo {year} {2018})}\BibitemShut
  {NoStop}%
\bibitem [{\citenamefont {Kl\"upfel}\ \emph {et~al.}(2009)\citenamefont
  {Kl\"upfel}, \citenamefont {Reinhard}, \citenamefont {B\"urvenich},\ and\
  \citenamefont {Maruhn}}]{klu09}%
  \BibitemOpen
  \bibfield  {author} {\bibinfo {author} {\bibfnamefont {P.}~\bibnamefont
  {Kl\"upfel}}, \bibinfo {author} {\bibfnamefont {P.-G.}\ \bibnamefont
  {Reinhard}}, \bibinfo {author} {\bibfnamefont {T.~J.}\ \bibnamefont
  {B\"urvenich}},\ and\ \bibinfo {author} {\bibfnamefont {J.~A.}\ \bibnamefont
  {Maruhn}},\ }\bibfield  {title} {\bibinfo {title} {Variations on a theme by
  {S}kyrme: A systematic study of adjustments of model parameters},\ }\href
  {https://doi.org/10.1103/PhysRevC.79.034310} {\bibfield  {journal} {\bibinfo
  {journal} {Phys. Rev. C}\ }\textbf {\bibinfo {volume} {79}},\ \bibinfo
  {pages} {034310} (\bibinfo {year} {2009})}\BibitemShut {NoStop}%
\bibitem [{\citenamefont {Chabanat}\ \emph {et~al.}(1997)\citenamefont
  {Chabanat}, \citenamefont {Bonche}, \citenamefont {Haensel}, \citenamefont
  {Meyer},\ and\ \citenamefont {Schaeffer}}]{cha97}%
  \BibitemOpen
  \bibfield  {author} {\bibinfo {author} {\bibfnamefont {E.}~\bibnamefont
  {Chabanat}}, \bibinfo {author} {\bibfnamefont {P.}~\bibnamefont {Bonche}},
  \bibinfo {author} {\bibfnamefont {P.}~\bibnamefont {Haensel}}, \bibinfo
  {author} {\bibfnamefont {J.}~\bibnamefont {Meyer}},\ and\ \bibinfo {author}
  {\bibfnamefont {R.}~\bibnamefont {Schaeffer}},\ }\bibfield  {title} {\bibinfo
  {title} {A {S}kyrme parametrization from subnuclear to neutron star
  densities},\ }\href
  {https://doi.org/https://doi.org/10.1016/S0375-9474(97)00596-4} {\bibfield
  {journal} {\bibinfo  {journal} {Nucl. Phys. A}\ }\textbf {\bibinfo {volume}
  {627}},\ \bibinfo {pages} {710} (\bibinfo {year} {1997})}\BibitemShut
  {NoStop}%
\bibitem [{\citenamefont {Poltoratska}\ \emph {et~al.}(2014)\citenamefont
  {Poltoratska}, \citenamefont {Fearick}, \citenamefont {Krumbholz},
  \citenamefont {Litvinova}, \citenamefont {Matsubara}, \citenamefont {von
  Neumann-Cosel}, \citenamefont {Ponomarev}, \citenamefont {Richter},\ and\
  \citenamefont {Tamii}}]{pol14}%
  \BibitemOpen
  \bibfield  {author} {\bibinfo {author} {\bibfnamefont {I.}~\bibnamefont
  {Poltoratska}}, \bibinfo {author} {\bibfnamefont {R.~W.}\ \bibnamefont
  {Fearick}}, \bibinfo {author} {\bibfnamefont {A.~M.}\ \bibnamefont
  {Krumbholz}}, \bibinfo {author} {\bibfnamefont {E.}~\bibnamefont
  {Litvinova}}, \bibinfo {author} {\bibfnamefont {H.}~\bibnamefont
  {Matsubara}}, \bibinfo {author} {\bibfnamefont {P.}~\bibnamefont {von
  Neumann-Cosel}}, \bibinfo {author} {\bibfnamefont {V.~Y.}\ \bibnamefont
  {Ponomarev}}, \bibinfo {author} {\bibfnamefont {A.}~\bibnamefont {Richter}},\
  and\ \bibinfo {author} {\bibfnamefont {A.}~\bibnamefont {Tamii}},\ }\bibfield
   {title} {\bibinfo {title} {Fine structure of the isovector giant dipole
  resonance in $^{208}\mathrm{Pb}$: Characteristic scales and level
  densities},\ }\href {https://doi.org/10.1103/PhysRevC.89.054322} {\bibfield
  {journal} {\bibinfo  {journal} {Phys. Rev. C}\ }\textbf {\bibinfo {volume}
  {89}},\ \bibinfo {pages} {054322} (\bibinfo {year} {2014})}\BibitemShut
  {NoStop}%
\bibitem [{\citenamefont {Carter}\ \emph {et~al.}(2022)\citenamefont {Carter}
  \emph {et~al.}}]{car22}%
  \BibitemOpen
  \bibfield  {author} {\bibinfo {author} {\bibfnamefont {J.}~\bibnamefont
  {Carter}} \emph {et~al.},\ }\bibfield  {title} {\bibinfo {title} {Damping of
  the isovector giant dipole resonance in $^{40,48}${C}a},\ }\href
  {https://doi.org/https://doi.org/10.1016/j.physletb.2022.137374} {\bibfield
  {journal} {\bibinfo  {journal} {Phys. Lett. B}\ }\textbf {\bibinfo {volume}
  {833}},\ \bibinfo {pages} {137374} (\bibinfo {year} {2022})}\BibitemShut
  {NoStop}%
\bibitem [{\citenamefont {von Neumann-Cosel}\ \emph {et~al.}(2019)\citenamefont
  {von Neumann-Cosel}, \citenamefont {Ponomarev}, \citenamefont {Richter},\
  and\ \citenamefont {Wambach}}]{vnc19b}%
  \BibitemOpen
  \bibfield  {author} {\bibinfo {author} {\bibfnamefont {P.}~\bibnamefont {von
  Neumann-Cosel}}, \bibinfo {author} {\bibfnamefont {V.~Y.}\ \bibnamefont
  {Ponomarev}}, \bibinfo {author} {\bibfnamefont {A.}~\bibnamefont {Richter}},\
  and\ \bibinfo {author} {\bibfnamefont {J.}~\bibnamefont {Wambach}},\
  }\bibfield  {title} {\bibinfo {title} {Gross, intermediate and fine structure
  of nuclear giant resonances: Evidence for doorway states},\ }\href
  {https://doi.org/10.1140/epja/i2019-12795-1} {\bibfield  {journal} {\bibinfo
  {journal} {Eur. Phys. J. A}\ }\textbf {\bibinfo {volume} {55}},\ \bibinfo
  {pages} {224} (\bibinfo {year} {2019})}\BibitemShut {NoStop}%
\bibitem [{\citenamefont {Efros}\ \emph {et~al.}(2007)\citenamefont {Efros},
  \citenamefont {Leidemann}, \citenamefont {Orlandini},\ and\ \citenamefont
  {Barnea}}]{efros2007}%
  \BibitemOpen
  \bibfield  {author} {\bibinfo {author} {\bibfnamefont {V.~D.}\ \bibnamefont
  {Efros}}, \bibinfo {author} {\bibfnamefont {W.}~\bibnamefont {Leidemann}},
  \bibinfo {author} {\bibfnamefont {G.}~\bibnamefont {Orlandini}},\ and\
  \bibinfo {author} {\bibfnamefont {N.}~\bibnamefont {Barnea}},\ }\bibfield
  {title} {\bibinfo {title} {The lorentz integral transform (lit) method and
  its applications to perturbation-induced reactions},\ }\href
  {https://doi.org/10.1088/0954-3899/34/12/R02} {\bibfield  {journal} {\bibinfo
   {journal} {J. Phys. G: Nucl. Part. Phys.}\ }\textbf {\bibinfo {volume}
  {34}},\ \bibinfo {pages} {R459} (\bibinfo {year} {2007})}\BibitemShut
  {NoStop}%
\bibitem [{\citenamefont {Nevo~Dinur}\ \emph {et~al.}(2014)\citenamefont
  {Nevo~Dinur}, \citenamefont {Ji}, \citenamefont {Bacca},\ and\ \citenamefont
  {Barnea}}]{NevoDinur:2014ngu}%
  \BibitemOpen
  \bibfield  {author} {\bibinfo {author} {\bibfnamefont {N.}~\bibnamefont
  {Nevo~Dinur}}, \bibinfo {author} {\bibfnamefont {C.}~\bibnamefont {Ji}},
  \bibinfo {author} {\bibfnamefont {S.}~\bibnamefont {Bacca}},\ and\ \bibinfo
  {author} {\bibfnamefont {N.}~\bibnamefont {Barnea}},\ }\bibfield  {title}
  {\bibinfo {title} {{Efficient method for evaluating energy-dependent sum
  rules}},\ }\href {https://doi.org/10.1103/PhysRevC.89.064317} {\bibfield
  {journal} {\bibinfo  {journal} {Phys. Rev. C}\ }\textbf {\bibinfo {volume}
  {89}},\ \bibinfo {pages} {064317} (\bibinfo {year} {2014})},\ \Eprint
  {https://arxiv.org/abs/1403.7651} {arXiv:1403.7651 [nucl-th]} \BibitemShut
  {NoStop}%
\bibitem [{\citenamefont {Hagen}\ \emph {et~al.}(2014)\citenamefont {Hagen},
  \citenamefont {Papenbrock}, \citenamefont {Hjorth-Jensen},\ and\
  \citenamefont {Dean}}]{hagen2014}%
  \BibitemOpen
  \bibfield  {author} {\bibinfo {author} {\bibfnamefont {G.}~\bibnamefont
  {Hagen}}, \bibinfo {author} {\bibfnamefont {T.}~\bibnamefont {Papenbrock}},
  \bibinfo {author} {\bibfnamefont {M.}~\bibnamefont {Hjorth-Jensen}},\ and\
  \bibinfo {author} {\bibfnamefont {D.~J.}\ \bibnamefont {Dean}},\ }\bibfield
  {title} {\bibinfo {title} {Coupled-cluster computations of atomic nuclei},\
  }\href {https://doi.org/10.1088/0034-4885/77/9/096302} {\bibfield  {journal}
  {\bibinfo  {journal} {Reports on Progress in Physics}\ }\textbf {\bibinfo
  {volume} {77}},\ \bibinfo {pages} {096302} (\bibinfo {year}
  {2014})}\BibitemShut {NoStop}%
\bibitem [{\citenamefont {Bacca}\ \emph {et~al.}(2013)\citenamefont {Bacca},
  \citenamefont {Barnea}, \citenamefont {Hagen}, \citenamefont {Orlandini},\
  and\ \citenamefont {Papenbrock}}]{bac13}%
  \BibitemOpen
  \bibfield  {author} {\bibinfo {author} {\bibfnamefont {S.}~\bibnamefont
  {Bacca}}, \bibinfo {author} {\bibfnamefont {N.}~\bibnamefont {Barnea}},
  \bibinfo {author} {\bibfnamefont {G.}~\bibnamefont {Hagen}}, \bibinfo
  {author} {\bibfnamefont {G.}~\bibnamefont {Orlandini}},\ and\ \bibinfo
  {author} {\bibfnamefont {T.}~\bibnamefont {Papenbrock}},\ }\href
  {https://doi.org/10.1103/PhysRevLett.111.122502} {\bibfield  {journal}
  {\bibinfo  {journal} {Phys. Rev. Lett.}\ }\textbf {\bibinfo {volume} {111}},\
  \bibinfo {pages} {122502} (\bibinfo {year} {2013})}\BibitemShut {NoStop}%
\bibitem [{\citenamefont {Miorelli}\ \emph {et~al.}(2018)\citenamefont
  {Miorelli}, \citenamefont {Bacca}, \citenamefont {Hagen},\ and\ \citenamefont
  {Papenbrock}}]{mio18}%
  \BibitemOpen
  \bibfield  {author} {\bibinfo {author} {\bibfnamefont {M.}~\bibnamefont
  {Miorelli}}, \bibinfo {author} {\bibfnamefont {S.}~\bibnamefont {Bacca}},
  \bibinfo {author} {\bibfnamefont {G.}~\bibnamefont {Hagen}},\ and\ \bibinfo
  {author} {\bibfnamefont {T.}~\bibnamefont {Papenbrock}},\ }\bibfield  {title}
  {\bibinfo {title} {Computing the dipole polarizability of $^{48}\mathrm{Ca}$
  with increased precision},\ }\href
  {https://doi.org/10.1103/PhysRevC.98.014324} {\bibfield  {journal} {\bibinfo
  {journal} {Phys. Rev. C}\ }\textbf {\bibinfo {volume} {98}},\ \bibinfo
  {pages} {014324} (\bibinfo {year} {2018})}\BibitemShut {NoStop}%
\bibitem [{\citenamefont {Jiang}\ \emph {et~al.}(2020)\citenamefont {Jiang},
  \citenamefont {Ekstr\"om}, \citenamefont {Forss\'en}, \citenamefont {Hagen},
  \citenamefont {Jansen},\ and\ \citenamefont {Papenbrock}}]{jiang20}%
  \BibitemOpen
  \bibfield  {author} {\bibinfo {author} {\bibfnamefont {W.~G.}\ \bibnamefont
  {Jiang}}, \bibinfo {author} {\bibfnamefont {A.}~\bibnamefont {Ekstr\"om}},
  \bibinfo {author} {\bibfnamefont {C.}~\bibnamefont {Forss\'en}}, \bibinfo
  {author} {\bibfnamefont {G.}~\bibnamefont {Hagen}}, \bibinfo {author}
  {\bibfnamefont {G.~R.}\ \bibnamefont {Jansen}},\ and\ \bibinfo {author}
  {\bibfnamefont {T.}~\bibnamefont {Papenbrock}},\ }\bibfield  {title}
  {\bibinfo {title} {Accurate bulk properties of nuclei from $a=2$ to
  $\ensuremath{\infty}$ from potentials with $\mathrm{\ensuremath{\Delta}}$
  isobars},\ }\href {https://doi.org/10.1103/PhysRevC.102.054301} {\bibfield
  {journal} {\bibinfo  {journal} {Phys. Rev. C}\ }\textbf {\bibinfo {volume}
  {102}},\ \bibinfo {pages} {054301} (\bibinfo {year} {2020})}\BibitemShut
  {NoStop}%
\bibitem [{\citenamefont {Simonis}\ \emph {et~al.}(2017)\citenamefont
  {Simonis}, \citenamefont {Stroberg}, \citenamefont {Hebeler}, \citenamefont
  {Holt},\ and\ \citenamefont {Schwenk}}]{sim17}%
  \BibitemOpen
  \bibfield  {author} {\bibinfo {author} {\bibfnamefont {J.}~\bibnamefont
  {Simonis}}, \bibinfo {author} {\bibfnamefont {S.~R.}\ \bibnamefont
  {Stroberg}}, \bibinfo {author} {\bibfnamefont {K.}~\bibnamefont {Hebeler}},
  \bibinfo {author} {\bibfnamefont {J.~D.}\ \bibnamefont {Holt}},\ and\
  \bibinfo {author} {\bibfnamefont {A.}~\bibnamefont {Schwenk}},\ }\bibfield
  {title} {\bibinfo {title} {Saturation with chiral interactions and
  consequences for finite nuclei},\ }\href
  {https://doi.org/10.1103/PhysRevC.96.014303} {\bibfield  {journal} {\bibinfo
  {journal} {Phys. Rev. C}\ }\textbf {\bibinfo {volume} {96}},\ \bibinfo
  {pages} {014303} (\bibinfo {year} {2017})}\BibitemShut {NoStop}%
\bibitem [{\citenamefont {Jansen}\ \emph {et~al.}(2011)\citenamefont {Jansen},
  \citenamefont {Hjorth-Jensen}, \citenamefont {Hagen},\ and\ \citenamefont
  {Papenbrock}}]{jansen2011}%
  \BibitemOpen
  \bibfield  {author} {\bibinfo {author} {\bibfnamefont {G.~R.}\ \bibnamefont
  {Jansen}}, \bibinfo {author} {\bibfnamefont {M.}~\bibnamefont
  {Hjorth-Jensen}}, \bibinfo {author} {\bibfnamefont {G.}~\bibnamefont
  {Hagen}},\ and\ \bibinfo {author} {\bibfnamefont {T.}~\bibnamefont
  {Papenbrock}},\ }\bibfield  {title} {\bibinfo {title} {Toward open-shell
  nuclei with coupled-cluster theory},\ }\href
  {https://doi.org/10.1103/PhysRevC.83.054306} {\bibfield  {journal} {\bibinfo
  {journal} {Phys. Rev. C}\ }\textbf {\bibinfo {volume} {83}},\ \bibinfo
  {pages} {054306} (\bibinfo {year} {2011})}\BibitemShut {NoStop}%
\bibitem [{\citenamefont {Jansen}(2013)}]{jansen2013}%
  \BibitemOpen
  \bibfield  {author} {\bibinfo {author} {\bibfnamefont {G.~R.}\ \bibnamefont
  {Jansen}},\ }\bibfield  {title} {\bibinfo {title} {Spherical coupled-cluster
  theory for open-shell nuclei},\ }\href
  {https://doi.org/10.1103/PhysRevC.88.024305} {\bibfield  {journal} {\bibinfo
  {journal} {Phys. Rev. C}\ }\textbf {\bibinfo {volume} {88}},\ \bibinfo
  {pages} {024305} (\bibinfo {year} {2013})}\BibitemShut {NoStop}%
\bibitem [{\citenamefont {Novario}\ \emph {et~al.}(2020)\citenamefont
  {Novario}, \citenamefont {Hagen}, \citenamefont {Jansen},\ and\ \citenamefont
  {Papenbrock}}]{novario2020}%
  \BibitemOpen
  \bibfield  {author} {\bibinfo {author} {\bibfnamefont {S.~J.}\ \bibnamefont
  {Novario}}, \bibinfo {author} {\bibfnamefont {G.}~\bibnamefont {Hagen}},
  \bibinfo {author} {\bibfnamefont {G.~R.}\ \bibnamefont {Jansen}},\ and\
  \bibinfo {author} {\bibfnamefont {T.}~\bibnamefont {Papenbrock}},\ }\bibfield
   {title} {\bibinfo {title} {Charge radii of exotic neon and magnesium
  isotopes},\ }\href {https://doi.org/10.1103/PhysRevC.102.051303} {\bibfield
  {journal} {\bibinfo  {journal} {Phys. Rev. C}\ }\textbf {\bibinfo {volume}
  {102}},\ \bibinfo {pages} {051303} (\bibinfo {year} {2020})}\BibitemShut
  {NoStop}%
\bibitem [{\citenamefont {Hagen}\ \emph {et~al.}(2022)\citenamefont {Hagen},
  \citenamefont {Novario}, \citenamefont {Sun}, \citenamefont {Papenbrock},
  \citenamefont {Jansen}, \citenamefont {Lietz}, \citenamefont {Duguet},\ and\
  \citenamefont {Tichai}}]{hagen2024}%
  \BibitemOpen
  \bibfield  {author} {\bibinfo {author} {\bibfnamefont {G.}~\bibnamefont
  {Hagen}}, \bibinfo {author} {\bibfnamefont {S.~J.}\ \bibnamefont {Novario}},
  \bibinfo {author} {\bibfnamefont {Z.~H.}\ \bibnamefont {Sun}}, \bibinfo
  {author} {\bibfnamefont {T.}~\bibnamefont {Papenbrock}}, \bibinfo {author}
  {\bibfnamefont {G.~R.}\ \bibnamefont {Jansen}}, \bibinfo {author}
  {\bibfnamefont {J.~G.}\ \bibnamefont {Lietz}}, \bibinfo {author}
  {\bibfnamefont {T.}~\bibnamefont {Duguet}},\ and\ \bibinfo {author}
  {\bibfnamefont {A.}~\bibnamefont {Tichai}},\ }\bibfield  {title} {\bibinfo
  {title} {Angular-momentum projection in coupled-cluster theory: Structure of
  $^{34}\mathrm{Mg}$},\ }\href {https://doi.org/10.1103/PhysRevC.105.064311}
  {\bibfield  {journal} {\bibinfo  {journal} {Phys. Rev. C}\ }\textbf {\bibinfo
  {volume} {105}},\ \bibinfo {pages} {064311} (\bibinfo {year}
  {2022})}\BibitemShut {NoStop}%
\bibitem [{\citenamefont {Sun}\ \emph {et~al.}(2024)\citenamefont {Sun},
  \citenamefont {Ekstr\"om}, \citenamefont {Forss\'en}, \citenamefont {Hagen},
  \citenamefont {Jansen},\ and\ \citenamefont {Papenbrock}}]{sun2024}%
  \BibitemOpen
  \bibfield  {author} {\bibinfo {author} {\bibfnamefont {Z.~H.}\ \bibnamefont
  {Sun}}, \bibinfo {author} {\bibfnamefont {A.}~\bibnamefont {Ekstr\"om}},
  \bibinfo {author} {\bibfnamefont {C.}~\bibnamefont {Forss\'en}}, \bibinfo
  {author} {\bibfnamefont {G.}~\bibnamefont {Hagen}}, \bibinfo {author}
  {\bibfnamefont {G.~R.}\ \bibnamefont {Jansen}},\ and\ \bibinfo {author}
  {\bibfnamefont {T.}~\bibnamefont {Papenbrock}},\ }\bibfield  {title}
  {\bibinfo {title} {{Multiscale physics of atomic nuclei from first
  principles}}\ }(\bibinfo {year} {2024})\ \Eprint
  {https://arxiv.org/abs/2404.00058} {arXiv:2404.00058 [nucl-th]} \BibitemShut
  {NoStop}%
\bibitem [{\citenamefont {Hu}\ \emph {et~al.}(2024)\citenamefont {Hu},
  \citenamefont {Sun}, \citenamefont {Hagen},\ and\ \citenamefont
  {Papenbrock}}]{hu2024}%
  \BibitemOpen
  \bibfield  {author} {\bibinfo {author} {\bibfnamefont {B.~S.}\ \bibnamefont
  {Hu}}, \bibinfo {author} {\bibfnamefont {Z.~H.}\ \bibnamefont {Sun}},
  \bibinfo {author} {\bibfnamefont {G.}~\bibnamefont {Hagen}},\ and\ \bibinfo
  {author} {\bibfnamefont {T.}~\bibnamefont {Papenbrock}},\ }\bibfield  {title}
  {\bibinfo {title} {Ab initio computations of strongly deformed nuclei near
  $^{80}\mathrm{Zr}$},\ }\href {https://doi.org/10.1103/PhysRevC.110.L011302}
  {\bibfield  {journal} {\bibinfo  {journal} {Phys. Rev. C}\ }\textbf {\bibinfo
  {volume} {110}},\ \bibinfo {pages} {L011302} (\bibinfo {year}
  {2024})}\BibitemShut {NoStop}%
\end{thebibliography}%

\end{document}